\documentclass[epj,preprint]{svjour}
\onecolumn
\usepackage{latexsym}
\usepackage{graphics}
\usepackage{comment}
\usepackage{amsmath,amssymb,bm}
\usepackage{multirow}
\usepackage{lineno}
\newcommand{\del}{$\Delta$}
\newcommand{\delpp}{$\Delta^{++}$}
\newcommand{\delp}{$\Delta^{+}$}

\newcommand{\deltodal}{$\Delta \to$Ne$^+$e$^-$}
\newcommand{\piztodal}{$\pi^0\to\gamma$e$^+$e$^-$}
\newcommand{\etatodal}{$\eta\to\gamma$e$^+$e$^-$}
\newcommand{\etatohad}{$\eta\to\pi^+\pi^-\pi^0$}
\newcommand{\pppnpip}{pp$\to$pn$\pi^+$}
\newcommand{\pppppiz}{pp$\to$pp$\pi^0$}
\newcommand{\ppppeta}{pp$\to$pp$\eta$}

\newcommand{\epem}{e$^+$e$^-$}

\newcommand{\piz}{$\pi^0$}
\newcommand{\pip}{$\pi^+$}
\newcommand{\pipp}{$\pi^+$p}
\newcommand{\pnpip}{pn$\pi^+$}
\newcommand{\pppiz}{pp$\pi^0$}

\newcommand{\pppippim}{pp$\pi^+ \pi^-$}
\newcommand{\pim}{$\pi^-$}
\newcommand{\gcsq}{GeV/c$^2$}
\newcommand{\gcsqsq}{GeV$^2$/c$^4$}

\title{ 
Study of 
exclusive one-pion and one-eta production using hadron and dielectron channels in pp reactions at kinetic beam energies of 1.25 GeV and 2.2 GeV with HADES}
\author{G.~Agakishiev$^{5}$, H.~Alvarez-Pol$^{15}$, A.~Balanda$^{2}$, R.~Bassini$^{10}$,
M.~B\"{o}hmer$^{8}$, H.~Bokemeyer$^{3}$, J.~L.~Boyard$^{13}$, P.~Cabanelas$^{15}$, S.~Chernenko$^{5}$, T.~Christ$^{8}$, M.~Destefanis$^{9}$, F.~Dohrmann$^{4}$, A.~Dybczak$^{2}$, T.~Eberl$^{8}$,
 L.~Fabbietti$^{7}$, O.~Fateev$^{5}$, P.~Finocchiaro$^{1}$, J.~Friese$^{8}$, I.~Fr\"{o}hlich$^{6}$, T.~Galatyuk$^{6,B}$, J.~A.~Garz\'{o}n$^{15}$, R.~Gernh\"{a}user$^{8}$,\\ C.~Gilardi$^{9}$, M.~Golubeva$^{11}$, D.~Gonz\'{a}lez-D\'{\i}az$^{C}$, F.~Guber$^{11}$, M.~Gumberidze$^{13}$, T.~Hennino$^{13}$, R.~Holzmann$^{3}$, A.~Ierusalimov$^{5}$, I.~Iori$^{10,E}$, A.~Ivashkin$^{11}$, M.~Jurkovic$^{8}$, B.~K\"{a}mpfer$^{4,D}$, K.~Kanaki$^{4}$, T.~Karavicheva$^{11}$, I.~Koenig$^{3}$, W.~Koenig$^{3}$, B.~W.~Kolb$^{3}$, 
 R.~Kotte$^{4}$, A.~Kozuch$^{2,F}$, F.~Krizek$^{14}$, W.~K\"{u}hn$^{9}$, A.~Kugler$^{14}$,
  A.~Kurepin$^{11}$, S.~Lang$^{3}$, K.~Lapidus$^{7}$, T.~Liu$^{13}$, L.~Maier$^{8}$, 
  J.~Markert$^{6}$, V.~Metag$^{9}$, B.~Michalska$^{2}$, E.~Morini\`{e}re$^{13}$, J.~Mousa$^{12}$, M.~M\"{u}nch$^{3}$, C.~M\"{u}ntz$^{6}$, L.~Naumann$^{4}$, J.~Otwinowski$^{2}$, Y.~C.~Pachmayer$^{6}$, V.~Pechenov$^{3}$, O.~Pechenova$^{6}$, T.~P\'{e}rez~Cavalcanti$^{9}$, J.~Pietraszko$^{6}$, V.~Posp\'{\i}sil$^{14}$, 
W.~Przygoda$^{2}$, B.~Ramstein$^{13}$, A.~Reshetin$^{11}$, M.~Roy-Stephan$^{13}$, A.~Rustamov$^{3}$,
A.~Sadovsky$^{11}$, B.~Sailer$^{8}$, P.~Salabura$^{2}$, M.~S\'{a}nchez$^{15}$, A.~Schmah$^{A}$,
E.~Schwab$^{3}$, Yu.G.~Sobolev$^{14}$, S.~Spataro$^{1,9,G}$, B.~Spruck$^{9}$, H.~Str\"{o}bele$^{6}$,
J.~Stroth$^{3,6}$, C.~Sturm$^{3}$, A.~Tarantola$^{6}$, K.~Teilab$^{6}$, P.~Tlusty$^{14}$, A.~Toia$^{9}$, M.~Traxler$^{3}$, R.~Trebacz$^{2}$, H.~Tsertos$^{12}$, V.~Wagner$^{14}$, M.~Wisniowski$^{2}$, 
T.~Wojcik$^{2}$, J.~W\"{u}stenfeld$^{4}$, S.~Yurevich$^{3}$,  Y.~Zanevsky$^{5}$, P.~Zumbruch$^{3}$ \thanks{\emph{Corresponding author:} ramstein@ipno.in2p3.fr}.}

\institute{
(HADES collaboration) \\\mbox{$^{1}$Istituto Nazionale di Fisica Nucleare - Laboratori Nazionali del Sud, 95125~Catania, Italy}\\
\mbox{$^{2}$Smoluchowski Institute of Physics, Jagiellonian University of Cracow, 30-059~Krak\'{o}w, Poland}\\
\mbox{$^{3}$GSI Helmholtz-Zentrum f\"{u}r Schwerionenforschung GmbH, 64291~Darmstadt, Germany}\\
\mbox{$^{4}$Institut f\"{u}r Strahlenphysik, Helmholtz-Zentrum  Dresden-Rossendorf, 01314~Dresden, Germany}\\
\mbox{$^{5}$Joint Institute of Nuclear Research, 141980~Dubna, Russia}\\
\mbox{$^{6}$Institut f\"{u}r Kernphysik, Goethe-Universit\"{a}t, 60438 ~Frankfurt, Germany}\\
\mbox{$^{7}$Excellence Cluster 'Origin and Structure of the Universe', 85748~Garching, Germany}\\
\mbox{$^{8}$Physik Department E12, Technische Universit\"{a}t M\"{u}nchen, 85748~Garching, Germany}\\
\mbox{$^{9}$II.Physikalisches Institut, Justus Liebig Universit\"{a}t Giessen, 35392~Giessen, Germany}\\
\mbox{$^{10}$Istituto Nazionale di Fisica Nucleare, Sezione di Milano, 20133~Milano, Italy}\\
\mbox{$^{11}$Institute for Nuclear Research, Russian Academy of Science, 117312~Moscow, Russia}\\
\mbox{$^{12}$Department of Physics, University of Cyprus, 1678~Nicosia, Cyprus}\\
\mbox{$^{13}$Institut de Physique Nucl\'{e}aire (UMR 8608), CNRS/IN2P3 - Universit\'{e} Paris Sud, F-91406~Orsay Cedex, France}\\
\mbox{$^{14}$Nuclear Physics Institute, Academy of Sciences of Czech Republic, 25068~Rez, Czech Republic}\\
\mbox{$^{15}$Departamento de F\'{\i}sica de Part\'{\i}culas, Univ. de Santiago de Compostela, 15706~Santiago de Compostela, Spain}\\
\\
\mbox{$^{A}$ also at Lawrence Berkeley National Laboratory, ~Berkeley, USA}\\
\mbox{$^{B}$ also at ExtreMe Matter Institute EMMI, 64291~Darmstadt, Germany}\\
\mbox{$^{C}$ also at Technische Universit\"{a}t Darmstadt, 64289~Darmstadt, Germany}\\
\mbox{$^{D}$ also at Technische Universit\"{a}t Dresden, 01062~Dresden, Germany}\\
\mbox{$^{E}$ also at Dipartimento di Fisica, Universit\`{a} di Milano, 20133~Milano, Italy}\\
\mbox{$^{F}$ also at Panstwowa Wyzsza Szkola Zawodowa, 33-300~Nowy Sacz, Poland}\\
\mbox{$^{G}$ also at Dipartimento di Fisica Generale and INFN, Universit\`{a} di Torino, 10125~Torino, Italy}\\
}

\abstract{We present measurements of exclusive $\pi^{+,0}$ and $\eta$ production 
in pp reactions at 1.25 GeV and 2.2 GeV beam kinetic energy in hadron and dielectron 
channels. In the case of \pip\ and \piz, high-statistics invariant-mass and 
angular distributions are obtained within the HADES acceptance
as well as acceptance corrected distributions, which are compared to a resonance model. 
The sensitivity of the data to the yield and production angular distribution 
of \del(1232) and higher lying baryon resonances is shown, and an improved parameterization 
is proposed. The extracted cross sections are of special interest 
in the case of \ppppeta\ , since controversial data exist at 2.0 GeV;
we find $\sigma =0.142 \pm 0.022$ mb. 
Using the dielectron channels, the \piz\ and $\eta$ Dalitz decay signals are 
reconstructed with yields fully consistent with the hadronic channels. 
The electron invariant masses and acceptance corrected  helicity angle 
distributions are found in good agreement with model predictions.  
}
\authorrunning{HADES collaboration}
\titlerunning{Exclusive one-pion and one-eta production with HADES}
\begin{document}
\maketitle
%

\section{Introduction}
\label{intro}
%
Meson production in nucleon-nucleon reactions in the kinetic beam energy range
1 -- 2 GeV tests an important sector of strong interaction on the hadron level.
 It is the subject of extensive studies by both experiment and theory
with the aim to establish eventually the link of hadron physics to QCD as fundamental theory. 
Most of the available   high-statistics data  originate from recent near-threshold measurements, at excess energies $<$ 150 MeV, performed at the SATURNE, CELSIUS and COSY facilities.
Phenomenological models, usually adjusted to data of these reactions, 
serve as step towards a concise theoretical description. Such models
are based on, for instance, the one-boson exchange (OBE) approximation
describing the production amplitudes by a coherent sum of meson, nucleon and 
baryon-resonance currents. They reveal that meson production is a complex process  
with important contributions from nucleon-nucleon final-state interactions and interferences 
between the contributing reaction channels even within a tree-level approach. 
This situation often leads to ambiguous 
model descriptions of the experimental results (for a review see \cite{Moskal_review02}).
For higher energies, i.e.\ excess energies $> 150$ MeV,
the data base is more scarce and originates mainly from 
low statistics bubble-chamber experiments \cite{Baldini88}.

 The production of $\pi$ and $\eta$ mesons in nucleon-nucleon collisions is of particular 
importance because of their coupling to baryonic resonances. Hence, experimental 
data on one-$\pi$ and one-$\eta$ production provide quantitative information on hadronic interactions, as well as resonance excitations and resonance properties.

Cross sections for pion production in the beam energy range from 0.6 to 1.5 GeV (excess energies between 140 and  500 MeV) have been provided in the past by  many  experiments  \cite{Bugg64,Eisner65,Bacon67,HudomaljGabitzsch78,Shimizu82,Wicklund87}. Meanwhile, also the precision of the measurements of differential distributions,  which is essential for unravelling the reaction mechanism,  has been improved 
\cite{Comptour94,Andreev94,Sarantsev04,ElSamad06,ElSamad09,Skorodko09,Ermakov11}. The dominance of the intermediate 
$\Delta(1232)$ production and the peripheral character of the reaction, which increases with energy, stand out very clearly
in this region. Comparison of the shape of various differential 
distributions from both exclusive reactions pp$\rightarrow$pp$\pi^0$ and 
pp$\rightarrow$pn$\pi^+$ in the range $0.6-0.94$ GeV to calculations within the one-pion exchange (OPE) show a nice agreement \cite{Andreev94,Sarantsev04}.  However, 
the magnitude of the cross sections for both reactions, regardless of the choice 
of the form of $\pi$-nucleon interaction, is explained by the models within 
an accuracy of 20 - 30\% only. This points to some missing elements in
the assumed reaction mechanism, 
as for example the exchange of heavier 
mesons, contributions of heavier resonances or/and non-resonant $\pi$ production, as demonstrated by 
 a recent detailed analysis of \pppnpip\ and \pppppiz\ reactions at a beam energy of $0.94$ GeV 
within the framework of a partial-wave analysis \cite{Ermakov11}.

 At  beam energies  higher than 1.5 GeV, the 
interpretation of the low-statistics bubble-chamber data \cite{Fickinger62,Alexander67,Coletti67}    was based on the isobar 
model assuming an incoherent sum of contributions from the decays of various baryon 
resonances into pions. However,  large uncertainties remain, due to the limited statistical significance of the 
corresponding experimental results. Further progress 
in the understanding of meson production in p+p interactions thus requires new high 
statistics data.

For the one-$\eta$ production, precise data have been collected close to the reaction 
threshold (excess energies below 120 MeV) 
\cite{chiavassa94,Chiavassa94,bergolt94,Hibou98,calen99,Roderburg00,Winter02,abdel03,cosy11,Smyrski00} and compared 
to various OBE models (for a survey cf.~\cite{Moskal_review02}). Most of the calculations indicate 
a dominant role of resonances, in particular the  $N^*(1535)$ formed via the exchange of 
virtual pseudoscalar ($\pi$, $\eta$) and  vector ($\rho,\omega$) mesons. However, 
the models differ in the description of how the resonance is excited, which is due to 
the uncertainty in the nucleon-meson-$N^*$ couplings. 
Better constraints  can be obtained from differential distributions and polarization observables, 
 as demonstrated in \cite{Czyz07}. 
The dominance of $N^*(1535)$ seems also to persist at higher beam energies ($2-3$ GeV), 
as shown by a detailed analysis of the $pp\eta$ Dalitz distributions by the DISTO 
collaboration \cite{Balestra04}. The latter work, however, does not provide 
absolute cross sections which are very important for the quantitative evaluation of 
the role of resonances.

Under the assumption that intermediate baryon resonances play a dominant role in 
$\pi$, $\eta$ and $\rho$  production, a model was developed \cite{Teis97}
based on an incoherent sum of various resonance  contributions. The matrix element of the  $\Delta(1232)$ production was calculated within the OPE model 
\cite{Dmitriev86}, which had been adjusted to available differential distributions of pion production in the \pppnpip\ channel at incident kinetic energies in the range  0.9-1.5 GeV. 
 The other matrix elements were kept constant and were determined by fitting the total meson production  cross sections.

Meson production is an important ingredient of microscopic transport models 
which were developed to describe heavy-ion collisions. Such approaches rely on a 
realistic treatment of elementary nucleon-nucleon and meson-nucleon interactions, 
as described in the previous paragraphs, to calculate double-differential cross sections 
of the produced particles \cite{Bleicher99,Shekhter03,Thomere07,Bratko08,Barz09,Schmidt09}.

Electromagnetic decays of mesons and baryon resonances are sources of \epem\ 
(dielectron) pairs which play a prominent role in heavy-ion physics as penetrating 
probes of nuclear and hadronic media. Therefore, data on $\pi$, $\eta$, and 
baryon resonance production in proton-proton interactions are essential for the 
understanding of \epem\  pair production in p+p, p+nucleus and nucleus-nucleus collisions. 
Recent precise measurements of the HADES collaboration underline the need to 
understand elementary sources of \epem\ pairs for the interpretation of 
heavy-ion data \cite{Agakichiev07_CC2GeV,Agakichiev08_CC1GeV,Agakichiev10_elem,Agakishiev_ArKCl2011}. 
It turns out that the baryon resonances are especially important. They contribute to 
the dielectron spectra via direct Dalitz decay, \deltodal ,
and via two-step processes in which the resonance decays 
into a nucleon and a meson with a subsequent 
\epem\ pair produced in the Dalitz decay of the meson ({\it e.g.}\  
$\Delta(1232) \to$N$\pi^0$  \textbf{•}
followed by $\pi^0 \to \gamma$\epem\ or  
N(1535)$\to$N $\eta$  followed by $\eta \to \gamma$\epem ) 
or two-body decays of the produced mesons 
({\it e.g.}\ N(1520)$\to$N$\rho$  followed by  $\rho \to$\epem ).

At higher beam energies \cite{Agakichiev11_pp35}, the dielectron yield is a 
complicated cocktail resulting from decays of many mesons and baryon resonances, 
but the main dielectron sources in nucleon-nucleon collisions are still the $\pi^0$ Dalitz 
decays for dielectron invariant masses M$_{ee} <$ 0.135 GeV/c${}^2$, the $\eta$  Dalitz 
decay contribution for 0.14 GeV/c${}^2$  $<$M$_{ee} < 0.547$ GeV/c${}^2$ and Dalitz decays of 
baryon resonances and light vector mesons ($\rho$ and $\omega$) for M$_{ee} > 0.6$ GeV/c${}^2$. 

In the past, experiments studying p+p and $\pi$+p interactions have either analyzed 
the hadronic or leptonic final states. The HADES apparatus \cite{Agakichiev09_techn}
allows  for the first time to measure hadron and 
\epem\ pair final states simultaneously with high statistics. With such data, it is 
possible to achieve a consistent description of meson production in p+p reactions in 
the hadron as well as in the dielectron channel. In this article, we present the first 
step in this direction and compare experimental results from the analysis of three 
reaction channels 
\pppnpip , 
\pppppiz ,  and 
\ppppeta\ 
measured at kinetic 
beam energies of 2.2 GeV and of the 
corresponding dielectron channels pp $\to$ pp$\pi^0 \to$ pp\epem$\gamma$ 
and pp$\to$pp$\eta \to$pp\epem$\gamma$ obtained at a kinetic 
beam energy of 2.2 GeV. At 1.25 GeV, which is below the threshold of $\eta$
production in $pp$ reactions,
only the reactions involving pions were analyzed. As reference model, we use predictions 
of the aforementioned resonance model of \cite{Teis97}, complemented with experimental 
results \cite{Balestra04}.  Differential spectra have been measured with high statistics 
providing strong constraints on the production mechanisms as well as on the different 
resonance contributions.

The polarization of virtual photons has never been measured in the Dalitz decays of the 
pseudoscalar mesons, although the distributions of the respective
 angles of the emitted lepton with respect to the virtual photon direction 
(the so called helicity) have been predicted for several sources \cite{Bratko95}. 
It  has also been suggested that such
distributions can  be used as important "fingerprints"  to distinguish between different 
dielectron sources in the inclusive measurements. Indeed, as shown in \cite{Arnaldi09_ang} 
, virtual photon polarization
appears to be a very important characteristics of the dielectron excess 
radiation originating from the hot and dense hadronic matter created in heavy-ion 
collisions at SPS energies.  In contrast to the situation at the SPS, similar 
measurements at SIS18 energies show a sizable anisotropy in the helicity distributions 
\cite{Agakishiev_ArKCl2011}. Therefore, it is desirable to determine the relevant distributions for 
the $\pi^0$ and $\eta$ mesons, as well as for baryon resonances, such as the $\Delta(1232)$, 
and bremsstrahlung in p+p collisions. In this work, we present such a measurement for 
the $\eta$ Dalitz decay which has been isolated in the 
pp$\to$pp\epem$\gamma$ reaction channel.

Our paper has the following structure.
The experimental set-up and event reconstruction will be briefly described in
sec.~\ref{sec|exp_aspects}.  The data analysis methods and simulation tools are 
presented in Secs.~\ref{sec:Data_analysis} and \ref{sec|res_model}. 
Section \ref{sec|results} is devoted to the
discussion of the results. We draw conclusions in sec.~\ref{sec|conclusion}.   \par

\section{Experimental aspects}
\label{sec|exp_aspects}
\subsection{Detector overview}
\label{sec|detector}
The High Acceptance Dielectron Spectrometer (HADES) consists of six identical sectors covering   polar angles  between 18$^{\circ}$ and 85$^{\circ}$ and between 65 and 90 $\%$ of the azimuthal range.  While a detailed description of the set-up can be found in \cite{Agakichiev09_techn},  we summarize here only the features  relevant  for the present analyses.   Proton beams with intensities up to 10$^7$ particles/s were directed to a  5~cm long liquid-hydrogen target  of 1$\%$ interaction probability. The  momenta of the produced particles are deduced from the hits in the four drift chamber planes (two before and two after the magnetic field zone) using  a Runge-Kutta algorithm \cite{Agakichiev09_techn}. The momentum resolution  is  2-3$\%$ for protons and  pions and 1-2$\%$ for electrons, depending on momentum and angle \cite{Agakichiev09_techn}. 
The trigger for the HADES experiments consists of two stages: The first level trigger (LVL1) is built on different configurations of hit multiplicity measurements in  two plastic scintillator walls for polar angles larger (TOF) and smaller (TOFINO)  than 45$^{\circ}$, respectively. The second level trigger (LVL2) selects  e$^\pm$ candidates defined by  a Ring Imaging Cerenkov detector (RICH) and information from TOF and an electromagnetic  shower detector (Pre-Shower) behind TOFINO. 
The analysis of hadronic channels  was based on   LVL1 triggered events selected by either of the two following configurations: The first required a coincidence between two hits in two opposite sectors of the time-of-flight detectors, with at least one in the TOFINO. This configuration was optimized  for the selection of pp elastic scattering events and also used for the \pppppiz\ and \pppnpip\ reactions. The second configuration was  based  on a charged particle multiplicity of four or more, at least one hit in the TOFINO and at least two signals in opposite sectors. This selection was used for the reconstruction of \ppppeta\ hadronic and  dielectron channels (\etatohad\ and \etatodal ) as well as  \pppppiz\ dielectron channels (\piztodal )  in 2.2 GeV collisions.  For the  \piz\ and  $\eta$ electromagnetic Dalitz decay measurements, in addition to the latter LVL1 configuration,  a LVL2 decision was requested, {\it i.e.} at least one electron candidate in the RICH.\par       

\subsection{Event reconstruction}
\label{sec|event_reconstruction}
Five different  final states were used for the study of exclusive $\pi$ and $\eta$ production  in pp reactions, as summarized in table \ref{tab|exit_channels}. An  important feature of the HADES apparatus, which is exploited in the present analysis, is the ability to measure  both hadrons  \cite{Agakichiev09_pionsCC,Agakichiev09_phi} and electrons \cite{Agakichiev07_CC2GeV,Agakichiev08_CC1GeV,Agakichiev10_elem} in the same experimental run.  Of special importance for the analysis of the dielectron channels is 
 the suppression of tracks  produced by photon conversion and consequently the reduction of  the combinatorial background \cite{Agakichiev09_techn}. This is achieved using criteria related to the track quality and the distance and opening angle between neighbouring tracks. Finally, only \epem\ pairs with  an opening angle  larger than 9$^0$  were propagated to the physics analysis. The  remaining combinatorial background is subtracted  from the measured unlike sign pair yields, using the arithmetical mean of the like sign  pair (e$^+$e$^+$ and e$^-$e$^-$) yield in the same event. 
\par  

\begin{table}[h!]
\centering
\begin{tabular}{|c|c|c|c|}
\hline
\multirow{2}{*}{reaction} & meson decay   &  measured & incident \\
 &  channel  & exit channel & energy \\
\hline
\pppnpip  &       - &\pipp (n) & 1.25, 2.2 GeV\\
\hline
\multirow{3}{*}{\pppppiz}  &  all & pp (\piz ) & 1.25, 2.2 GeV\\
\cline{2-4}
&  \piz $\to \gamma$\epem  & pp\epem ($\gamma$ ) &2.2 GeV\\
&  (BR 1.12$\%$) & &\\
\hline
\multirow{3}{*}{\ppppeta}& $\eta \to$ \pip\pim  \piz &    \pppippim (\piz ) & 2.2 GeV\\
& (BR 22.7$\%$) &    & \\
\cline{2-4}
& $\eta \to \gamma$\epem  &  pp\epem ($\gamma$) & 2.2 GeV\\
& (BR 0.68$\%$) &   & \\
\hline 
\end{tabular}
\caption{Investigated reaction channels for exclusive $\pi$ and $\eta$ production reactions.}
\label{tab|exit_channels}
\end{table}
\begin{figure}
	\centering
\resizebox{0.45\textwidth}{!}{
		\includegraphics{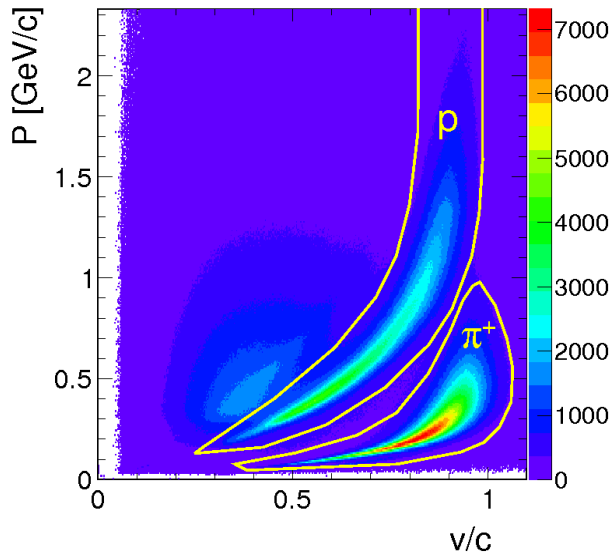}}
	\caption{(Color on-line) Correlation between momentum (p) and velocity (v/c) for particles without a RICH signal and with curvature corresponding to a positive charge. The overlaid yellow curves indicate graphical cuts to select protons and pions.}
	\label{fig:pid_nppiplus_22}
\end{figure}
  Since the RICH is hadron blind  in the given energy range and the   particle multiplicity is low, electrons and positrons are  selected using only the matching of a charged track reconstructed in the drift chambers and a ring pattern in the RICH detector.  Particle identification (PID) for pions and protons is provided  by the correlation between the velocity ($\beta$=v/c) obtained from TOF or TOFINO scintillator walls and the momentum deduced from the track deflection in the magnetic field \cite{Agakichiev09_techn}. The   start signal for the time measurements was taken from the fastest signal from the scintillator wall. To reconstruct the time-of-flight for each particle, a dedicated method was developed \cite{Agakichiev09_techn}, using the  identification of one reference particle,  the time-of-flight of which  can be calculated. When a \epem\ pair was present in the exit channel, one of the leptons could be used as the reference particle.
For events without electron candidates but containing a  negatively charged particle, it was used as a  reference particle and assigned the pion mass.   When only two positive  hadrons  were  observed in the final state, two hypotheses were tested: 1) the presence of two protons (2p events) and 2) the presence of  one pion and one proton (\pipp\ events). For each hypothesis, both hadrons could be used as reference particles, hence providing an additional consistency check.  In all the cases described above, the time-of-flight of the reference particle was calculated, and the velocities of all the other products were then deduced, using only the time-of-flight differences to the reference particle.  The correlation between velocity  and momentum of all particles was then used to reject the wrong hypotheses and to assign  the final  PID of all particles. 
    Figure~\ref{fig:pid_nppiplus_22} displays  such a correlation for positively charged tracks  without  signal in the RICH.  The efficiency of the PID procedure was higher than 90$\%$ for both pions and protons. In addition, in the case of the four-particle exit channels, the algorithm was checked in a dedicated experiment with a low beam intensity  using a START detector, as discussed in \cite{Agakichiev09_techn}.\par
\subsection{Acceptance and efficiency considerations}
\label{sec|acc_res}
 The spectrometer acceptance, detector efficiency and resolution  as well as the analysis cuts necessary to extract the signal   introduce important constraints  on the determination of the cross sections and on the comparison of the experimental distributions to model predictions.  For the HADES case, to compensate acceptance  losses due to spectrometer  geometry and extract 4$\pi$ integrated yields,  extrapolation  into the unmeasured regions of phase space is usually achieved  by means of a model. The  reliability of the model to describe the  shape of the relevant distributions also in the unmeasured regions determines the systematic errors of the acceptance corrections. \par
On the other hand, direct comparisons of theoretical and experimental distributions can  also be made inside the HADES acceptance using dedicated filters. For this purpose, the acceptances and efficiencies for the different particles (i.e electrons, pions and protons)  were separately tabulated in matrices as a function of momentum, azimuthal and polar angles. The matrix coefficients   have been determined using full GEANT simulations, with  all reaction products processed through the detector, and analysed with the same programmes  as done for real events. The resulting acceptance matrices describe the HADES fiducial volume only and can be applied as a filter to events generated  by models.  The corresponding efficiency  matrices account for the detection and reconstruction process and  have been used to correct the experimental data event-by-event.  In addition, an emulator of the trigger condition was  applied both on experimental data and simulated events. In the case of the two-hit trigger,  however, the data were corrected for the condition of having at least one particle in the TOFINO. 
The detection and reconstruction efficiency was  typically  90$\%$  for protons  and  pions and about 50$\%$  for electrons. In addition, the yields measured in the \epem\ channels were corrected for the LVL2 efficiency. The latter  was calculated by comparing the pp$\to$pp\epem X yield in unbiased LVL1 events to the yield obtained with both LVL1 and LVL2  conditions; in this way a LVL2 efficiency of 90$\pm$5\% was obtained. 
\par
The momentum resolution parameters were determined from the simulations  in bins  of momentum and polar and azimuthal angles and rescaled to match the resolution determined experimentally using the  elastic pp scattering \cite{Agakichiev09_techn}.
 The acceptance  matrices and resolution parameters,  necessary to filter and smear the model generated particle distributions before  comparing to HADES efficiency corrected data,  are available from the authors upon request. 

\subsection{Normalization procedure using elastic scattering}
\label{sec|elastic}
\begin{figure}
\begin{center}
\resizebox{0.45\textwidth}{!}{
  \includegraphics{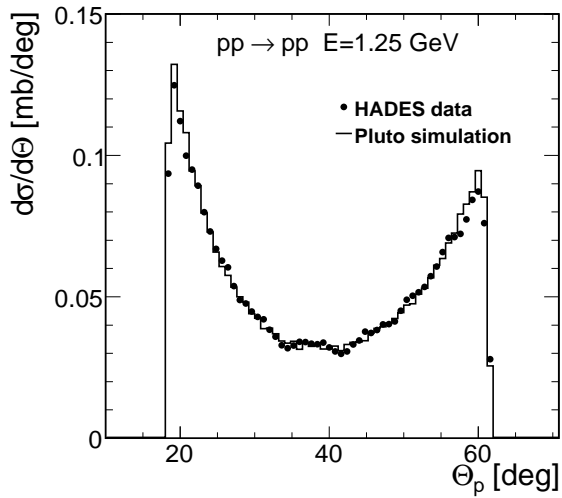}}
  \caption{Angular distribution in the laboratory system of measured pp elastic events (full dots) compared to PLUTO simulations (full histogram) using the angular distributions of \cite{Albers04}  as input. }
	\label{fig:elastic_ang_dist}
\end{center}
\end{figure}
The normalization of the experimental yield is obtained using the analysis of events produced in elastic scattering. 
Momentum conservation in the pp elastic scattering reaction leads to the two following relations between the polar angles $\theta_1$ and $\theta_2$ and azimuthal angles $\phi_1$ and $\phi_2$ of both protons: 
\begin{eqnarray}
\left|\phi_1-\phi_2\right|&=&180^0 ,
 \label{eq|corrang1}\\
 \tan\theta_1\tan\theta_2&=&\frac{1}{\gamma_{CM}^2},
 \label{eq|corrang2}
\end{eqnarray}
where $\gamma_{CM}$ is the Lorentz factor of the center-of-mass system.  The elastic events were  selected by an elliptic cut in the ($|\phi_1-\phi_2|$, $\tan\theta_1\tan\theta_2$) plane,  with semi axes corresponding to approximately 3$\sigma$ for each variable, {\it i.e.} $\pm$2.4$^{\circ}$ for $\left|\phi_1-\phi_2\right|$ and $\sigma$= 0.027 for $\tan\theta_1\tan\theta_2$.  The  angular distributions of the resulting event ensemble is corrected for efficiency and  compared to a simulation which uses the high precision data from the EDDA experiment \cite{Albers04} as input, as shown in fig.~\ref{fig:elastic_ang_dist} for the 1.25 GeV incident energy case. The histogram in fig.~\ref{fig:elastic_ang_dist} represents the angular distribution of events generated by PLUTO \cite{Froehlich07,Froehlich09} according to the parameterization of EDDA data and subjected to the HADES filter, as  described in more details in sec.~\ref{sec|acc_res}. The shape of the angular distribution is well reproduced, demonstrating that the angular dependence of the efficiency correction is under control.  The cut for angles larger than 62$^{\circ}$ reflects the  cut on the
 forward partner at about 18$^{\circ}$, which is due to the detector
 acceptance. The  experimental yield  was scaled in order to reproduce the simulated yield inside the HADES acceptance.  The resulting factors are  used for the normalization of the differential cross sections and have a precision of  about 6$\%$ at 1.25 GeV and 11$\%$ at 2.2 GeV, reflecting mainly the uncertainty on the global efficiency of the reconstruction and analysis.       \par
\section{Data analysis}
\label{sec:Data_analysis}
\subsection{Hadronic channels}
\subsubsection{Selection of  \pppppiz\ and \pppnpip\ channels }
\label{sec|pionselect}
\begin{figure}
	\centering
\resizebox{0.45\textwidth}{!}{		\includegraphics{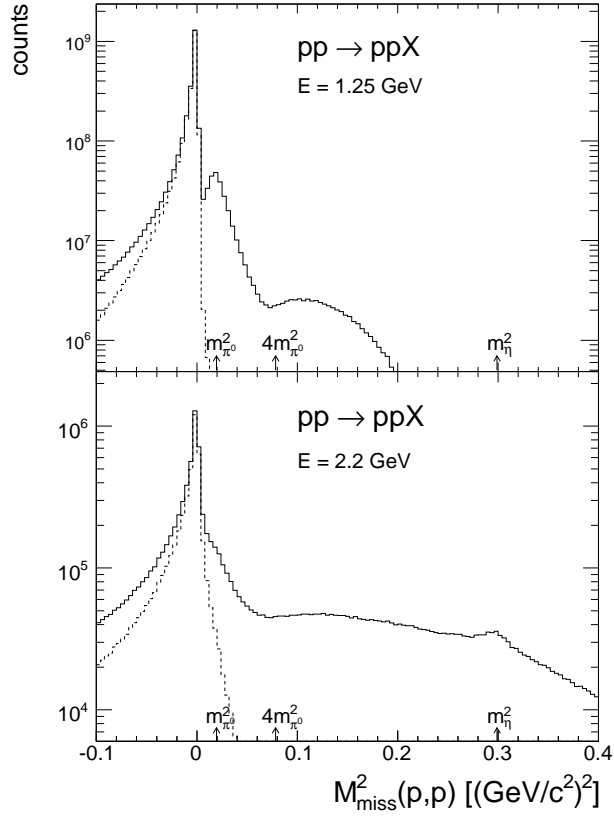}}
	\caption{Squared missing mass spectra for the reaction pp$\to$ppX at 1.25 GeV (top) and 2.2 GeV (bottom). The  dashed histogram shows the  events selected by the elastic scattering angular correlations eqs.(\ref{eq|corrang1},\ref{eq|corrang2}).}
		\label{fig:mm2p_2energies}
\end{figure}
To study the \pppppiz\ and \pppnpip\ channels, only events with two protons (2p) or one proton and one \pip\ (p\pip ) have been considered, respectively \cite{Wisnia09,Liu10}.
The selection of both channels is based on the requirement
 for the missing mass  to the system of the two detected charged particles to be close to the missing neutral particle mass. 
For the events with two detected protons,  the distribution of the squared missing mass to the two-proton system  (M$_{miss}^2$(p,p)), shown in  fig.~\ref{fig:mm2p_2energies}, present for both energies  a  prominent asymetric peak close to zero.  This contribution, clearly due to the elastic scattering, nicely  fulfills  the corresponding angular correlation (see sec.~\ref{sec|elastic}), as shown by the dashed histogram.    The different widths of these peaks at both energies as well as the better separation of the one-pion contribution at 1.25 GeV result from the resolution on the proton momentum. In addition, these spectra  reflect the larger phase space available for inelastic processes at 2.2 GeV. The  contribution of the  2$\pi$ contribution, visible for M$_{miss}^2$(p,p)  larger than 0.08 (GeV/c$^2$)$^2$, is clearly enhanced and  the $\eta$ meson production shows up for M$_{miss}^2$(p,p) around 0.3 (GeV/c$^2$)$^2$.  

\begin{figure}
	\centering
\resizebox{0.45\textwidth}{!}{
		\includegraphics{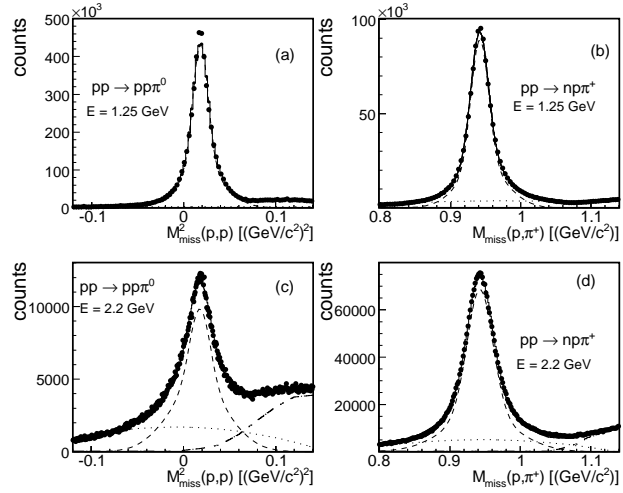}}
	\caption{Selection of events from \pppnpip\ and \pppppiz\ reactions based on a fit of missing masses. In the case of 2.2 GeV incident energy, the 2$\pi$  production deduced from the simulation  (dash-dotted curves) is subtracted before the fit. The fitting function consists of a sum of one polynomial (dotted curves) and  two-Gaussians (the sum of which is shown as dashed curves). The solid curves show  the sums of all the contributions.}
	\label{fig:mm_fits}
	\end{figure}
	To proceed with the selection of the exclusive one-pion production channels, the elastic events, selected as explained above, were first removed from the 2p sample.  The resulting  M$_{miss}^2$(p,p) spectra, shown in fig.~\ref{fig:mm_fits}a and~\ref{fig:mm_fits}c, are peaked close to the squared pion mass (m$_{\pi}^2$=0.02 \gcsqsq ). In the case of the p\pip\  events, the distributions of the missing mass to the p\pip\  system (M$_{miss}$(p,\pip )) are shown in fig.~\ref{fig:mm_fits}b, and the unmeasured neutrons become visible as peaks around 0.94 \gcsq . 	 The contribution of two-pion production is seen on the right hand side of the peaks. 
At 2.2 GeV,  this channel was simulated  as resulting from a double \del\ production, with normalization adjusted such as to fit the data at the highest missing masses, as shown in fig.~\ref{fig:mm_fits}c and~\ref{fig:mm_fits}d. For each  phase space bin considered in the analysis, the two-pion contribution was subtracted and  the remaining yield was then fitted with a function consisting of the sum of  two gaussians  plus a polynomial background. The signal was defined as the yield above this background; systematic errors were of the order of 5$\%$. 
For the 1.25 GeV data set, the two-pion production region  was simply excluded from the fit. Systematic errors for the signal yield   were estimated  from a variation of the background parameterization to be of the order of 1-3$\%$.
\subsubsection{Selection of the pp$\to$pp$\eta \to$ pp\pip \pim \piz\ channel}
\label{sec|etahad}
\begin{figure}[h!]
	\centering	
\resizebox{0.45\textwidth}{!}{\includegraphics{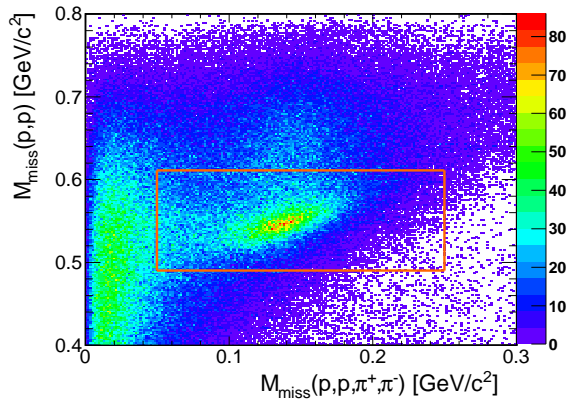}}
	\caption{(Color on-line) Analysis of the  \pppippim\ events in the pp reaction at 2.2 GeV. The correlation between the  missing mass to the two-proton system (M$_{miss}$(p,p)) and the missing mass to the four-particle  (M$_{miss}$(p,p,\pip ,\pim )) is shown. Events from the pp$\to$pp\pip\pim\ reaction are visible for M$_{miss}$(p,p,\pip ,\pim ) close to zero. The $\eta$ peak is clearly visible for M$_{miss}$(p,p) around the $\eta$ mass and M$_{miss}^{pp\pi^+\pi^-}$ close to the $\pi^0$ mass. The orange rectangle indicates the region used to further extract the $\eta$ signal.}
		\label{fig:mm_eta}
\end{figure}
To investigate the pp$\to$ pp$\eta \to$ pp\pip \pim \piz\ channel, events with two protons, one positive and one negative pion (pp\pip \pim events) are considered \cite{Perez06,Rustamov06,Spataro06}. Two observables have been defined: the missing masses  M$_{miss}$(p,p) and M$_{miss}$(p,p,\pip ,\pim ) to the  two-proton and  four-particle systems, respectively. The correlation between these two missing masses is displayed in fig.~\ref{fig:mm_eta}. The concentration of events with a  missing mass to the \pppippim\ system slightly above zero is due to the pp$\to$pp\pip\pim\ reaction; the broad structure with  M$_{miss}$(p,p,\pip ,\pim ) around the \piz\ mass corresponds to the pp$\to$\pppippim\piz\ final state and contains an elongated spot around M$_{miss}$(p,p)=0.55 \gcsq,  clearly due to the  \ppppeta\ signal. 
To extract the latter, first a selection of  M$_{miss}$(p,p,\pip ,\pim ) between 0.05 and 0.25 \gcsq , (corresponding to the vertical edges of the rectangle in fig.~\ref{fig:mm_eta}), was applied in order to reject most of the \pppippim\ background.\par 
\begin{figure}[h!]
	\centering
\resizebox{0.45\textwidth}{!}{		\includegraphics{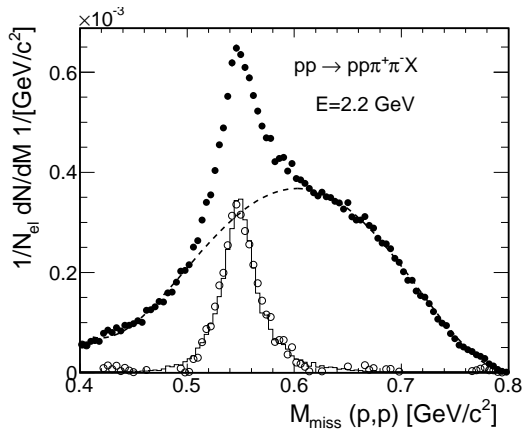}}
	\caption{Distribution of the missing mass to the two-proton  system (M$_{miss}$(p,p)) for the \pppippim\ events measured in the pp reaction  at 2.2 GeV, after selection on the  missing mass to the four-particle syatem (M$_{miss}$(p,p,\pip ,\pim)) around 
	the \piz\ mass (full black dots). The data are normalized to the pp elastic yield. The dashed curve shows the fit of the  non resonant  three-pion background. The  empty circles  result from  the subtraction of this background and define the $\eta$ signal. The  full histogram is the result of the simulation of the pp$\to$pp$\eta$ reaction.}
		\label{fig:signal_eta}
	 \end{figure}
	 The resulting (M$_{miss}$(p,p) spectrum, normalized to the elastic yield is  displayed in fig.~\ref{fig:signal_eta} and shows a peak at the mass of the $\eta$ meson on top of a broad continuum, which is mainly due to the non-resonant \pip\pim\piz\ production.  Its contribution in the  peak region (i.e. missing masses  between 0.490 and 0.610 GeV/c$^2$, delimited by the horizontal edges of the rectangle in fig.~\ref{fig:mm_eta})  was obtained from a polynomial fit of the data outside the peak region.
	 The $\eta$ signal was defined as the yield above this background, corresponding to about 24800 counts. 
	 The sensitivity to the background suppression was studied by  varying the limits for the fit. It  gave a systematic error of the order of $\pm$4$\%$. The missing mass distribution obtained from the simulation of the pp$\to$pp$\eta$ channel is  shown  as a full histogram  in  fig.~\ref{fig:signal_eta}. Its width  depends only marginally on the ingredients of the model for the $\eta$ production, which will be discussed in more details in   sec.~\ref{sec|res_model}. The  agreement of simulation and  experimental signal confirms  the consistency   of the extracted  $\eta$ signal and the good description of the detector resolution in the simulation.

\subsection{Exclusive dielectron channels at 2.2 GeV beam energy pp$\mathbf{\to}$pp$\mathbf{\pi^0}$/$\mathbf{\eta \to}$ ppe$\mathbf{^+}$e$\mathbf{^-\gamma}$}
\label{sec|pi0etadal_signal}

 \begin{figure}[h]
	\centering
 \resizebox{0.48\textwidth}{!}{     		\includegraphics{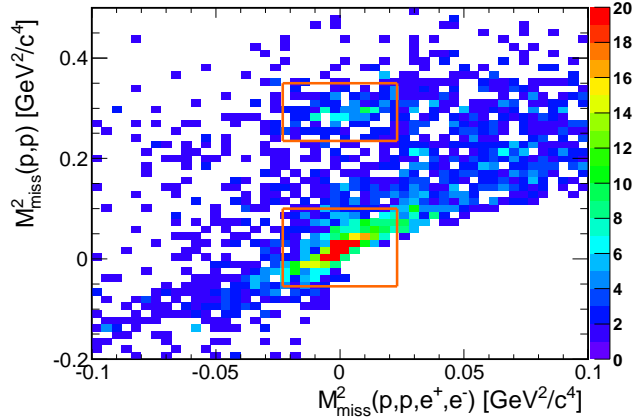}}
             \caption{(Color on-line) Analysis of the pp\epem\ events in the pp reaction at 2.2 GeV. Correlation between the square of the missing mass to the two-proton system  (M$_{miss}^2$(p,p)) and the square of the  missing mass to the four-particle system  (M$_{miss}^2$(p,p,e$^+$,e$^-$)) (see text). Events from the pp$\to$pp\epem $\gamma$ reaction are visible for M$_{miss}^2$(p,p,e$^+$,e$^-$) around zero. The orange rectangles show the limits used to extract the \piz\ and $\eta$ Dalitz decay signals.}
\label{fig:mmppee_bidim}
\end{figure}
 \begin{figure}[h]
	\centering
 \resizebox{0.48\textwidth}{!}{     		\includegraphics{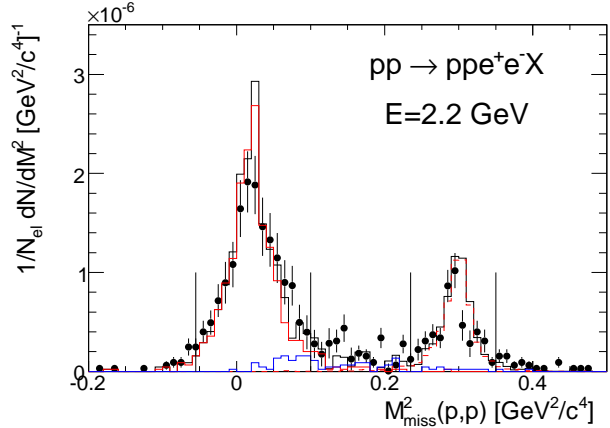}}
            \caption{(Color on-line) Distribution of squared missing mass  to the  two-proton system M$_{miss}^2$(p,p) (black dots) for pp\epem\ events after a cut on the  missing mass to the four-particle system ($\left| \mathrm{M}_{miss}^2 \mathrm{(p,p,e^+,e^-)} \right| <$ 0.023 \gcsqsq ). The yields are divided by the pp elastic scattering yield. The histograms show the results of GEANT simulations.  The red histograms peaking at the \piz\ and $\eta$ squared masses correspond to the exclusive \piz\ and $\eta$ production, respectively, followed by Dalitz decay, with cross sections as listed in table \ref{tab|res_channels}. The blue  histogram shows the contribution of multipion background, which is subtracted from the data, as explained in the text. The vertical lines depict the limits used to extract the experimental signals strength.}  \rm 
\label{fig:mmppee}
\end{figure}

To reconstruct the \pppppiz /$\eta \to$ pp\epem$\gamma$ channels at 2.2 GeV beam energy,  all events with two protons and one dielectron  (pp\epem )  have been  selected, utilizing  conversion pair rejection, as described in sec.~\ref{sec|event_reconstruction} and \cite{Wisnia09,Rustamov06,Spataro06,Spruck06}.  
 The same analysis procedure was also applied to events containing like-sign pairs (ppe$^+$e$^+$ and ppe$^-$e$^-$). The combinatorial background, defined as the arithmetical mean of the corresponding distributions for the latter two event samples, were then subtracted from the unlike-sign pair distributions. 
The signal-to-background  ratios are of the order of 3 in the \piz\ region and of 4  in the $\eta$ region.
In the same way as for the pp\pip \pim\ events,  the missing masses to the two-proton system  M$_{miss}$(p,p) and to the four-particle system  M$_{miss}$(p,p,e$^+$,e$^-$) were reconstructed, respectively.\par 
 The correlation between the squares of both missing masses  is shown in fig.~\ref{fig:mmppee_bidim} after combinatorial background subtraction.  Events from the pp$\to$pp\epem $\gamma$ reaction are visible for M$_{miss}^2$(p,p,e$^{+}$,e$^{-}$) around zero. 
  The contributions from  pp$\to$pp\piz /$\eta$  reactions followed by Dalitz decays \piz /$\eta \to$\epem$\gamma$  can be seen for M$_{miss}^2$(p,p) close to the \piz\ and  $\eta$ squared masses, respectively. The regions where both signals are extracted are shown  as rectangles  in fig.~\ref{fig:mmppee_bidim}. 
 The main remaining background is due to \epem\ pairs from the Dalitz decay of a \piz\ produced  in multi-pion production processes. Its  contribution,  of the order of 5$\%$  has been simulated using the cross sections 1.09 mb for \piz \piz\ production and 0.50 mb for the \piz \piz \piz  and \piz\pip\pim\  \cite{Baldini88}, and is removed bin by bin, as illustrated in fig.~\ref{fig:mmppee}.   Systematic errors  of the order of 3$\%$ and 8$\%$ have been estimated for the \piz\ and $\eta$ signal yields, respectively, by varying the missing mass limits and the shape of the multipion background. \par
A total amount of 6800 $\pm$ 82$_{stat}$   \piztodal\ events and 235 $\pm$ 19$_{stat}$  \etatodal\ events have been extracted.  The dashed  histogram in fig.~\ref{fig:mmppee}  is the result of the simulation of \piz\ and $\eta$ Dalitz decays, with ingredients based on a resonance model, as will be explained  in the following. The widths of the missing mass peaks, which do not depend on the details of the model and mainly reflect the momentum resolution of the particle tracks, are similar to the experimental ones.  In the same way as for the other analysis channels, this gives confidence on the reliability of  the procedure to extract the proper signal selection cuts from the simulation.

\section{Simulation of the reaction channels}  
%
\label{sec|res_model}
%
\begin{table}[H!]
\centering
\begin{tabular}{|c|c|c|c|c|}
\hline
 \multirow{2}{*}{final state} & intermediate  &\multicolumn{2}{|c|}{$\sigma _{\mathrm{Teis}}$(mb) \cite{Teis97}} &$\sigma_{\mathrm {adj}}$ (mb)  \\ 
\cline{3-5}
  &      process             &  1.25 GeV& 2.2 GeV &   2.2 GeV  \\ \hline
   pp  &pp  elastic& 23.5   & 17.8  &-\\ \hline
\multirow{6}{*}{pn\pip } & pp$\to$\del $^{++}$(1232) n      & 16.80  &    10.80 &  10.80\\
  \cline{2-5}
                        & pp$\to$ \del $^+$(1232) p                & 1.87  &   1.20 & 1.20 \\ 
  \cline{2-5}
                         & pp$\to$N$^{\star}$(1440) p        & 0.30    & 0.82  &  1.60\\
  \cline{2-5}
                         & pp$\to$N$^{\star}$(1520) p        & 0   & 0.18  &  0.36 \\
   \cline{2-5}                         & pp$\to$N$^{\star}$(1535) p        & 0   & 0.19  & 0.64 \\
   \cline{2-5}                      
                   &   non resonant &     0            &  0  &  0.30\\
  \cline{2-5} 
                            & total         & 18.97     &   13.09  & 14.90  \\
   \hline   
\multirow{3}{*}{pp\piz}  &   pp$\to$ \del $^+$(1232) p         & 3.73   & 2.40  &  2.40   \\                 \cline{2-5}
                          & pp$\to$ N$^{\star}$(1440) p      & 0.15  &   0.41 & 0.80\\                 
  \cline{2-5}
                          & pp$\to$ N$^{\star}$(1520) p      & 0  &   0.09  & 0.18\\                 
 \cline{2-5}              & pp$\to$ N$^{\star}$(1535) p      & 0  &   0.10  & 0.32\\                 
 \cline{2-5}                   &   non resonant &     0             &  0  & 0.15\\
  \cline{2-5} 
                          &  total               &  3.88                 &   2.99   & 3.85\\ 
 \hline 
\multirow{3}{*}{ pp$\to$pp$\eta$}  & pp$\to$ N$^{\star}$(1535) p  & 0  & 0.0725& 0.082 \\ 
  \cline{2-5}  
                   &   non resonant &     0             &  0.0525  & 0.060\\
     \cline{2-5}
    & total  &  0             &  0.125        &  0.142     \\\hline  
\end{tabular}
\caption{Cross sections used in the simulation. Elastic pp cross sections taken from \cite{Albers04} and \cite{PDG10} are used for the normalisation of the measurements.   For the inelastic channels, the first set of cross sections ($\sigma _{\mathrm{Teis}}$) is taken from \cite{Teis97} and is used in model A (see sec.\ref{sec|res_model}) at both energies  and in model B (see sec.\ref{sec|modif}) at 1.25 GeV.  For the $\eta$ production,   the ratio of N$^{\star}$(1535) to non-resonant production is taken from DISTO \cite{Balestra04}. 
The second set ($\sigma _{\mathrm{adj}}$), used in model B at 2.2 GeV,  is adjusted to  the HADES data (see text). 
}
\label{tab|res_channels}
\end{table}
%

 The inputs of our simulation are inspired by the resonance model by Teis et al., which is the basis for  the coupled channel BUU  transport code (CBUU) \cite{Teis97}. \rm The contribution of \del (1232), which is dominant for the pion production at the lowest energies, is taken from   the One-Pion-Exchange (OPE) model of Dmitriev et al. \cite{Dmitriev86}, which describes quite well the measured  cross sections, invariant mass distributions, and angular distributions of the \pppnpip\ reaction at incident energies between 0.97 and 3.2 GeV \cite{Bugg64,Eisner65,Coletti67}.  One important parameter of the model, which had been fitted to reproduce these data, is the cut-off  parameter $\Lambda_{\pi}$=0.63 GeV, entering the $\pi$N$\Delta$ and  $\pi NN$ vertex form factor $$F(t)=\frac{\Lambda_{\pi}^2-m_{\pi}^2}{\Lambda_{\pi}^2-t}$$, with t being the four-momentum transfer squared and m$_{\pi}$ the pion mass.  The available cross section values for exclusive  one-pion, two-pion or $\eta$ production in pp and pn reactions were used to fit the contributions of isospin 1/2 
(N$^{\star}$) and isospin 3/2 resonances other than the \del (1232) \cite{Teis97}. \par
For the simulation of the  channels analysed in our experiment, we employed the
 event generator PLUTO \cite{Froehlich07} and included   in its data base the  cross sections for the different reactions   (see table \ref{tab|res_channels}). The cross sections in the two first columns are directly taken from  \cite{Teis97} and the ones in the last columns are adjusted to better describe the present data, as will be shown in sec.~\ref{sec|results}.  The following relations 
 derived from the isospin coefficients, are fulfilled in the simulation:
 \begin{eqnarray}
 \sigma (\mathrm{pp} \to \Delta^{++} \mathrm{n}\to \pi^+ \mathrm{p n}) & = &9 \sigma (\mathrm{pp} \to \Delta^{+} \mathrm{p}\to \pi^+ \mathrm{n p})\\
 &=& \frac{9}{2}  \sigma (\mathrm{pp} \to \Delta^{+} \mathrm{p}\to \pi^0 \mathrm{p p})\\
\mathrm{and}\ \sigma (\mathrm{pp} \to \Delta \mathrm{N}\to \pi^+ \mathrm{p n}) & = &5 \sigma (\mathrm{pp} \to \Delta^{+} \mathrm{p}\to \pi^0 \mathrm{p p}). 
   \label{eq|isospin} 
 \end{eqnarray}
 In the same way, one gets  for the I=1/2 resonances,
 $$\sigma (\mathrm{pp} \to \mathrm{N}^{\star} \mathrm{p} \to \pi^+ \mathrm{n p})=  2  \sigma (\mathrm{pp} \to  \mathrm{N}^{\star} \mathrm{p} \to \pi^0 \mathrm{p p}).$$
    Resonances heavier than N(1535), which,  in the original Teis fit \cite{Teis97}, contribute  7$\%$ and  11$\%$ to the   \pnpip\ and \pppiz\ final states, are  neglected in our approach.
 As described in more detail in \cite{Froehlich07}, the resonance mass distributions  were taken according to \cite{Teis97}. Besides the  already mentioned case of the \del (1232), the  angular distributions for the production of the other resonances are assumed to be isotropic in the pp center-of-mass frame, as in the original Teis model \cite{Teis97}, except   for the N(1440) resonance, where a steep distribution following the One-Boson-Exchange (OBE) model of \cite{Huber94} was implemented. The decay angular distributions were kept isotropic, as in \cite{Teis97}, except for the \del (1232).  In this case, the angular distribution of the \del\ decay (\del $\to$N$\pi$) behaves as 1+0.65 cos$^2\theta$,  where $\theta$ is the angle between the $\pi$ momentum in the \del\ rest frame and the momentum transfer calculated in the rest frame of the excited nucleon. Such a shape was indeed found to reproduce  the available data \cite{Bacon67,Wicklund87}.   \par
For the $\eta$ production, a non-resonant contribution was introduced, in addition to the N(1535) (see table \ref{tab|res_channels}) with the same  proportion  as in the analysis of the DISTO data \cite{Balestra04}, measured at similar beam energies, and was simulated following phase-space distributions. For both resonant and non-resonant contributions, the angular distribution  was deduced from the DISTO data. The $\eta$ production cross section, which was not  measured in the DISTO experiment, is taken from \cite{Teis97}. The description of the Dalitz decay of $\eta$ and \piz\ mesons  uses Vector Meson Dominance (VMD) model form factors (cf. \cite{Froehlich07}).\par
Finally, the pp and pn final state interactions 
 have been implemented using the Jost function to weight the distributions  in the simulation \cite{Titov00}.
This model, including the  changes with respect to the original resonance model \cite{Teis97} mentionned above and summarized in table~ \ref{Tab:modelAmodelB} is called ''model A''  in the following. 
In the course of the paper, a second model (''model B'')  will be introduced, which consists in a better parameterization of the measured data.
 \begin{table}
\centering
\begin{tabular}{|c|c|c|}
\hline  & model A & model B \\ 
\hline 1.25 GeV &  (1) & (1) and (2)  \\ 
\hline 2.2 GeV &  (1) &  (1) and (3) \\ 
\hline 
\end{tabular}
\caption{Summary of the modifications introduced in models A and B with respect to the resonance model \cite{Teis97} (see text for more details).  (1):  pp and pn Final State Interaction, anisotropic $\Delta$(1232) decay angular distribution, N(1440) production angular distribution from \cite{Huber94}. (2): $\Lambda_{\pi}$ cut-off parameter changed from 0.63 GeV to 0.75 GeV and $\Delta(1232)$ production angular distribution further adjusted to describe the neutron angular distribution in the pp$\rightarrow$pn$\pi^+$ channel. (3) change of production cross sections for N(1440), N(1520)  and  N(1535) resonances and introduction of a non-resonant contribution following table~\ref{tab|res_channels}.}
\label{Tab:modelAmodelB}
\end{table} 
%
\section{Results and comparison with resonance model}
\label{sec|results}
\subsection{Dominance of \del\ resonance in one-pion production channels}

\label{sec|delta}
\begin{figure}
	\centering
\resizebox{0.45\textwidth}{!}{
		\includegraphics{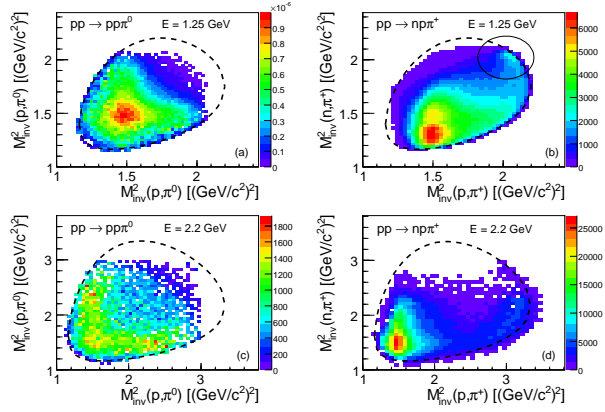}}
	\caption{(Color on-line) Dalitz plots of the pp$\to$\pppiz\ (left panels) and pp$\to$\pnpip\ (right panels) reactions: n\pip\  and p\piz\  invariant mass squared distributions at 1.25 GeV (upper row) and 2.2 GeV(lower row). In panel (c), the region affected by the final state interaction of the reaction  \pppnpip\  is marked by a circle.}
	\label{fig:Dalitz_polts}
\end{figure}
After selecting events from the one-pion production channels, following the procedure  described in sec.~ \ref{sec|pionselect},
we first investigate  the Dalitz plots (fig.~\ref{fig:Dalitz_polts}) with respect to the footprints of a resonant behaviour of particle production.  
For the \pppiz\ channel, an accumulation of yield for M$_{inv}^2$(p,\piz)=1.5 (GeV/c$^2$)$^2$, corresponding to the excitation of the \del$^{+}$ resonance is clearly seen at both incident energies.\par
For the \pppnpip\ reaction, the \del$^{++}$ signal stands out markedly at M$_{inv}^2$(p,\pip )=1.5 (GeV/c$^2$)$^2$, while the  \del$^{+}$ signal located at M$_{inv}^2$(n,\pip )=1.5 (GeV/c$^2$)$^2$ is not visible. 
The dashed curves in fig.~\ref{fig:Dalitz_polts} indicate the kinematical limits of the Dalitz plot for the different channels.   The empty zones in the plots are due to the acceptance 
cuts, the dominant effect being due to the  minimum proton  polar detection angle of about 18$^{\circ}$.
 For the \pppnpip\ reaction at 1.25 GeV, the  enhanced population for both M$_{inv}^2$(p,\pip ) and M$_{inv}^2$(n,\pip ) around 2 (GeV/c$^2$)$^2$ is due to the pn Final State Interaction (FSI), which enhances events with small relative momentum 
between the proton and the neutron. The FSI is less apparent  in the \pnpip\ channel at 2.2 GeV, since it  affects events with proton 
 angles below the acceptance limit. For the \pppiz\ channel, the pp FSI has a maximum effect when both protons hit the same sector
 of the HADES detector,  which is suppressed by the trigger configuration.
\begin{figure*}
	\centering
\resizebox{0.9\textwidth}{!}{\includegraphics{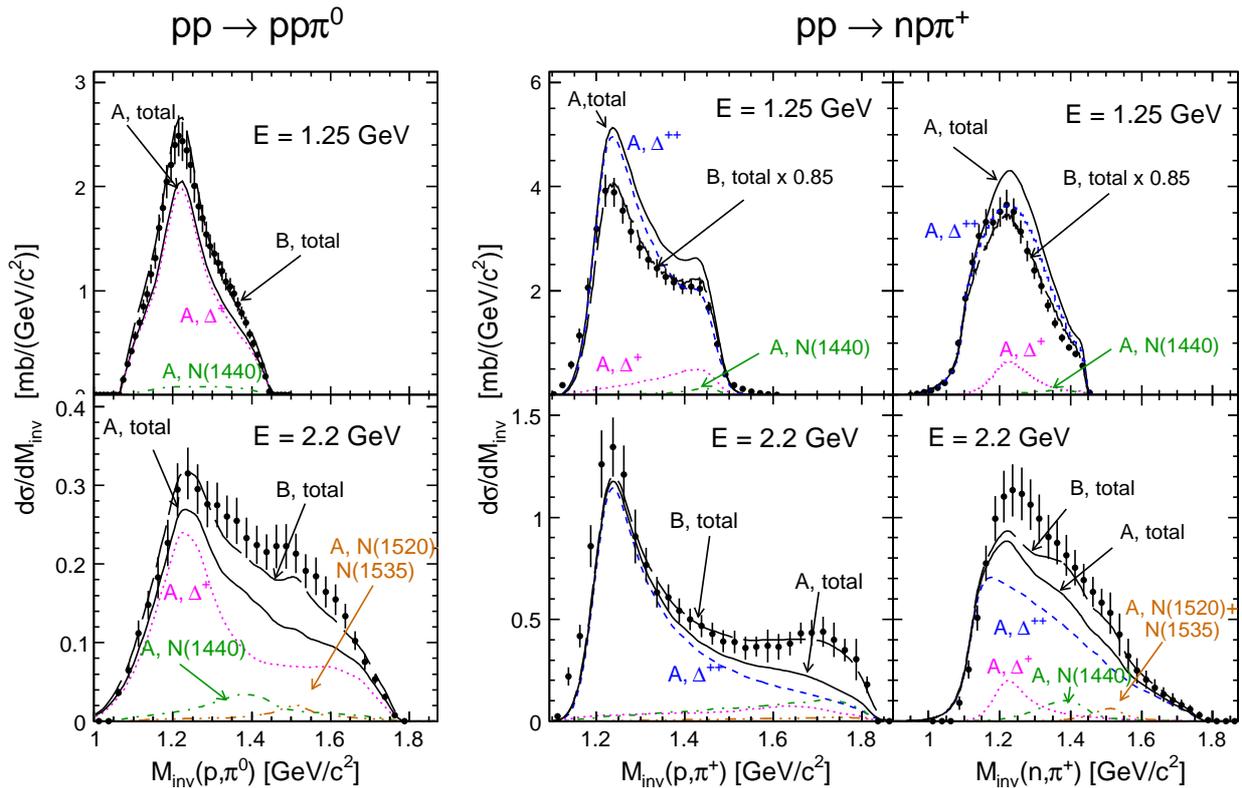}}
			\caption{(Color on-line) $\pi$N invariant mass distributions (full dots) measured in pp$\to$pp$\pi ^0$ and pp$\to$pn$\pi ^+$ reactions at 1.25 GeV (top row) and 2.2 GeV (bottom row). The data  are compared inside the detector acceptance on an absolute scale to the predictions of the  model A (see sec.~\ref{sec|res_model})  with contributions of  \del$^+$(1232) (pink dotted curve), \del$^{++}$(1232) (dashed blue curve), N(1440) (green short dash-dotted curve) and N(1520)+N(1535) (long dash-dotted light brown curve). The long dashed curve shows the result of the model B  with a scaling factor of 0.85 applied in the case of \pppnpip  at 1.25 GeV (see sec.~\ref{sec|modif}).}
	\label{fig:invariant_masses_small}
\end{figure*}

Figure~\ref{fig:invariant_masses_small}  exhibits respectively the p\piz\ invariant mass for the pp$\to$pp\piz\ reaction in the left part and the p\pip\ and n\pip\ invariant masses for the \pppnpip\ reaction in the   right part. The data  are corrected for reconstruction efficiencies and normalized using the total pp elastic cross
section, as explained in secs~\ref{sec|acc_res} and \ref{sec|elastic}. 
Error bars include statistical and systematic errors
due to signal  selection (1-5$\%$) and efficiency corrections (5-10$\%$). In addition, both isospin channels at a given energy are affected by the same global normalisation uncertainty of the order of 6$\%$ at 1.25 GeV and 11 $\%$ at 2.2 GeV.  The  M$_{inv}$(p, \pip ) and M$_{inv}$(p, \piz )  distributions are peaked around 
1.23 \gcsq , which confirms that most of the pions are produced via \del\ decay, although the distributions are obviously distorted by the acceptance. 
 The different contributions of the simulation with cross sections taken from \cite{Teis97} are shown in fig.~\ref{fig:invariant_masses_small}, too.  At both energies, the M$_{inv}$(p, \pip ) and M$_{inv}$(p, \piz ) distributions are mainly sensitive to the \del\  contributions.  The trend of the data is rather well reproduced, although obvious discrepancies concerning both the yields and the shapes can be observed.\par
  At 1.25 GeV, the model A overestimates the experimental yield by  20$\%$ for \pnpip\ and underestimates it by 20$\%$ for \pppiz . For the \pnpip\ channel, this discrepancy is  slightly larger than expected by taking into account, on the one hand, the discrepancies of the fit to previous data in both isospin channels  and, on the other hand, the combined uncertainty of normalisation (about $\pm$ 7 $\%$) and global efficiciency corrections (about $\pm$ 8 $\%$). However, the yields are obtained  here in a limited region of the phase space, and they are therefore sensitive to the distributions used in the model, as will be shown in the following. \par
    At 2.2 GeV, the contributions of higher lying resonances clearly show up at high invariant masses and are underestimated in the simulation using the cross sections from \cite{Teis97} (see table \ref{tab|res_channels}). \par

\subsection{Analysis of the pp$\mathbf{\to}$pn$\mathbf{\pi^{+}}$ channel}
We will now discuss in some detail the distributions obtained in the different channels. The results are first compared to the resonance model \cite{Teis97}  to show its capability to describe the data and then a better parameterization of the data is proposed. 
\subsubsection{\del\ resonance angular distributions}
\label{sec|angn}
\begin{figure}
	\centering 
        \resizebox{0.42\textwidth}{!}{
		\includegraphics{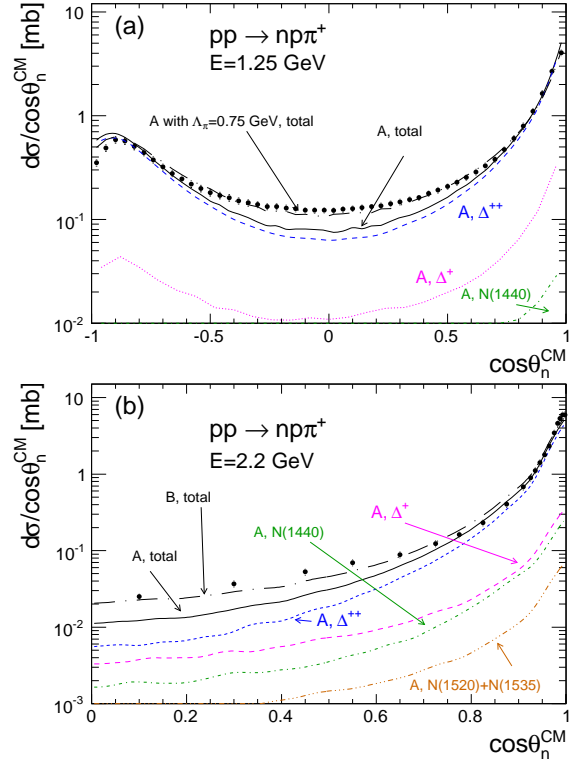}}
	\caption{(Color on-line) Angular distributions of the neutron in the pp center-of-mass system. Top: E=1.25 GeV, Bottom: E=2.2 GeV. Data (black points) are compared to
simulations with model A (solid curve) including \delpp (blue dashed curve), \delp\ (pink dotted curve), N$(1440)$ (green dash-dotted curve), N(1520) and N(1535) (brown dot-dot-dashed curve).  The dot-dashed curves show the result of model A with $\Lambda_{\pi}$=0.75 GeV in the case of 1.25 GeV and model B in the case of 2.2 GeV. Both calculations are scaled to reproduce the respective integrated experimental yield.}
	\label{fig:neutron_ang_hades}
\end{figure}
The neutron angular distributions in the center-of-mass system  measured in the \pppnpip\ channel at both energies are displayed in
 fig. \ref{fig:neutron_ang_hades}. These distributions mainly reflect the angular distribution of \del\ resonance 
production, since  pp$\to$n\delpp\ is the dominant process for the exclusive \pip\ production.  They  are strongly forward/backward 
 peaked, as expected for  the characteristic peripheral production of the \del\ resonance.  The distribution in fig.~\ref{fig:neutron_ang_hades}a (\pppnpip\ at 1.25 GeV) is highly distorted in the backward hemisphere, which is mainly due to  the limited acceptance for protons at small laboratory angles ($\theta <18 ^{\circ}$). At 2.2 GeV (fig.~\ref{fig:neutron_ang_hades}b), these acceptance losses are even larger. Therefore, we  did  not use the backward hemisphere  at this energy. We included into fig.~\ref{fig:neutron_ang_hades} the results from the simulations. In order to compare the shapes of the neutron angle distributions, the simulations were rescaled to reproduce the integrated experimental yields. At 1.25 GeV, it can be seen that  the forward/backward asymmetry is quite well reproduced by the simulation. The distribution is somewhat less peaked in the case of  the \delp\ and N$^{\star}$ excitations, since  the neutron comes from the decay of the resonance. The \delpp\ contribution  however still dominates by an order of magnitude  around cos $\theta_n$=0.  The slope at forward angles is well described by the sum of the different components (solid curve), but the experimental distribution is  slightly less steep than  the simulated one, at both energies. With the chosen normalisation to the integrated yield, 
the experimental yield at cos $\theta_n$=0 is larger than the simulated one by factors of about 1.6 and 2.5 at 1.25 GeV and  2.2 GeV, respectively. \par

\begin{figure}
	\begin{center}
	\resizebox{0.4\textwidth}{!}{
		\includegraphics{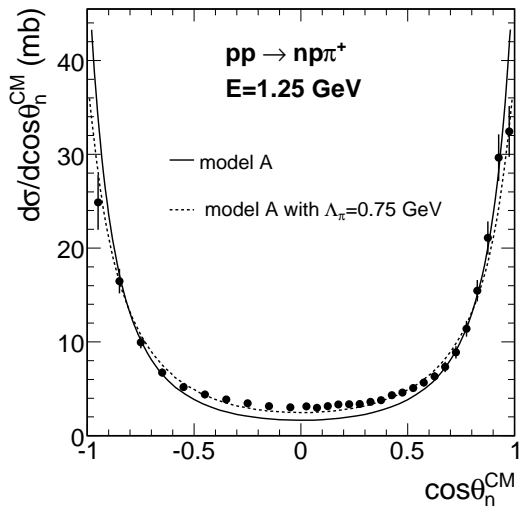}}
	\caption{Angular distribution of neutron in center-of-mass system after acceptance correction for the \pppnpip\  reaction at 1.25 GeV. Data (black points) are compared to simulations based on model A with $\Lambda_{\pi}$=0.63 GeV (solid curve) and the modified version  with $\Lambda_{\pi}$=0.75 GeV (dashed curve). Both simulation curves are normalized to reproduce the integrated experimental yield.}
	\label{fig:neutron_ang_AccCorr_125}
	\end{center}
	\end{figure}

Acceptance corrected angular distributions, which are useful to provide a result independent of the detector geometry, can only be obtained using a model.  
This could be  done with a good precision, at 1.25 GeV only, where the reaction mechanism  is best under control, due to the overwhelming contribution of the \del\ resonance. The acceptance correction factors are calculated for different (cos $\theta_n$,M$_{inv}$(p, \pip)) cells, chosen to optimize the precision of the correction and defined as the ratio of events from simulation in full phase space and in geometrical HADES acceptance.  In this way, the factors  depend weakly  on how the invariant mass and angular distributions of the \del\ are realized in the model. Remaining  uncertainties come mainly from the decay angle distribution of the \del\ resonance.
Due to the limited acceptance of our experiment, the measured distributions of the \pip\ emission angle do not allow  to improve the results from  previous measurements \cite{Bacon67,Wicklund87}, in which  a decay 
angle distribution compatible with 1+Bcos$^2\theta$  with  B=0.65 $\pm$0.30, was measured. 
The uncertainty on this anisotropy parameter has therefore been taken into account to calculate  the systematic errors.  The acceptance corrected neutron angular distribution  obtained in this way is shown with statistical and systematic errors in fig.~\ref{fig:neutron_ang_AccCorr_125}.  
Once corrected for acceptance, the neutron angular distribution recovers the expected forward/backward {symmetry.} The integral of this distribution gives the  cross section for the \pppnpip\ reaction, as will be discussed in sec.~\ref{sec|crossections}. The prediction from the resonance model is also shown, on fig.~\ref{fig:neutron_ang_AccCorr_125}, after a renormalization by a factor 0.85 to match the integrated yield of the experimental data.  The underestimation  of the experimental yield  around cos $\theta_n\approx 0$ is consistent with the result obtained within the HADES acceptance.\par
Since the shape of the \del\ production angular distribution in the OPE model depends on the value of the cut-off parameter $\Lambda_{\pi}$, the sensitivity of the simulation to this parameter was studied, keeping the cross sections of the different contributions as in \cite{Teis97}.  For $\Lambda_{\pi}$=0.75 GeV,  instead of the standard value of 0.63 GeV, the difference between model and experimental data at   cos $\theta_n$=0 is reduced from 40$\%$ to about 15$\%$ (see figs.~\ref{fig:neutron_ang_hades}a and ~\ref{fig:neutron_ang_AccCorr_125}). 
     It is clear, however, that the discrepancy of the measured angular distributions with respect to the OPE model might have a different origin than just a refitting of the cut-off parameter.  In the region of cos $\theta_n$=0, $\rho$ meson exchange could be more important due to the higher four-momentum transfer. In addition, non-resonant contributions  might have a much flatter angular distribution than the \del\ contribution. Finally, the interference between the amplitudes of the different resonances are neglected in our description. \par

\subsubsection{Modifications of the resonance model}
\label{sec|modif}
Considering the aforementioned deviations of the experimental distributions with respect to the resonance model of \cite{Teis97},  some modifications were introduced to provide a better parameterization of the data.\par
At 2.2 GeV, the \del\ resonance contributions were not changed, but the cross sections of higher lying resonances (N(1440), N(1520) and N(1535)) were increased and a non-resonant contribution, generated with a phase-space distribution, was added. The new  cross sections are listed in the last column of table \ref{tab|res_channels}. 
 A better description  of the  p\pip\ and \pip n invariant mass distributions in fig.~\ref{fig:invariant_masses_small} and of the neutron angular distribution in fig.~\ref{fig:neutron_ang_hades} in the pn\pip\ channel can be obtained, as can be seen by the long dash-dotted curves on the corresponding pictures. \par
At 1.25 GeV, we have chosen to keep the cross sections of \cite{Teis97}, but, in order to provide a parameterization of the data yet more precise than the OPE model with $\Lambda_{\pi}$=0.75 GeV (see sec.~\ref{sec|angn}), an iterative procedure was used to fit the \del\ production angular distribution such as to reproduce the measured neutron angular distribution. Due to the dominance  of the \delpp\ excitation in the \pppnpip\ reaction, this angular distribution is very close to the neutron angular distribution presented in fig.~\ref{fig:neutron_ang_AccCorr_125}. \par
 Applying an overall normalisation factor of 0.85, which is consistent with the different errors, (see  sec.~\ref{sec|crossections}), the yields and shapes of the invariant mass spectra  are well described by this modified resonance model, as can be seen by the long-dashed curve in fig.~\ref{fig:invariant_masses_small}.  Besides, the pn FSI,  introduced already in model A, while affecting only a very small fraction of the events, is found important to reproduce the behaviour of the distribution for the highest  invariant masses. 
 In this way, a new parameterization of the data is proposed at both energies, called "model B". The modifications with respect to model A are summarized in table \ref{Tab:modelAmodelB}. This parameterization will be checked for the \pppppiz\ channel in sec.~\ref{sec|pizero}, while in the next section, the invariant masses measured at 1.25 GeV are presented in more detail.

\subsubsection{Invariant mass distributions at 1.25 GeV}

\begin{figure}[h]
\begin{center}
\resizebox{0.49\textwidth}{!}{\includegraphics{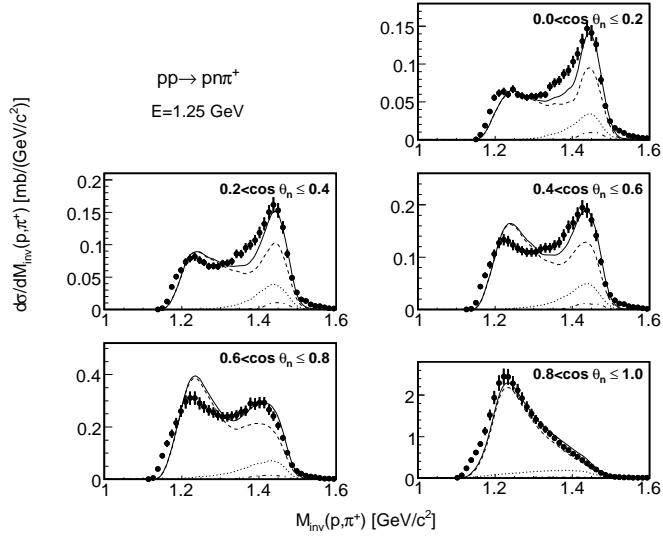}}
			\caption{ p\pip\ invariant mass distributions  measured in pp$\to$pn$\pi ^+$ reactions at 1.25 GeV for different bins in cos $\theta_n$ (full dots), compared to  model B (see text) rescaled by a factor 0.85, with total (full curves),   \delpp (1232) (dashed curves), \del$^{+}$(1232) (dotted curves), and N(1440) (dash-dotted curves) contributions.}
	\label{fig:invariant_masses_bins_pip}
	\end{center}
	\end{figure}
	\begin{figure}[h]
	\begin{center}
\resizebox{0.49\textwidth}{!}{\includegraphics{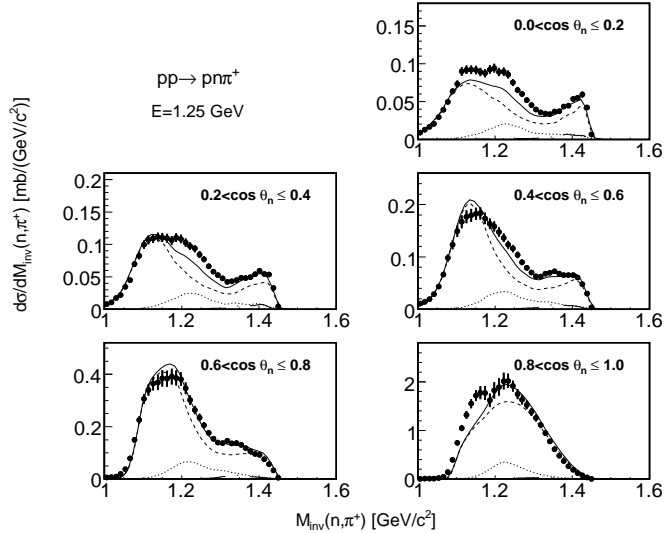}}
		\caption{ Same  as fig.~\ref{fig:invariant_masses_bins_pip}, but for for \pip n invariant mass distributions.}
		\label{fig:invariant_masses_bins_pin}
\end{center}
\end{figure} 
To further understand the contributing mechanisms in their impact  on the neutron angle cos $\theta_n$, the  p\pip \ and \pip n invariant mass spectra were  studied in the forward hemisphere in five different cos $\theta_n$ bins, as shown in figs.~\ref{fig:invariant_masses_bins_pip} and \ref{fig:invariant_masses_bins_pin}.  The spectra are compared to the results of the  simulation with model B, which takes into account  the experimental neutron angle distribution (see sec.~\ref{sec|modif}). As before, the model B is scaled by a factor 0.85 to reproduce the experimental yield integrated over the neutron angle.   The evolution of the shapes of the invariant mass spectra  as a function of neutron angle in the simulation is mainly due to the detector acceptance and trigger effects on the dominant \delpp\ contribution. In particular,  the structure at 
 about 1.45 GeV   for cos $\theta_n < 0.8$  is due to the requirement for the pion and proton to hit two opposite sectors. 
Although discrepancies of the order of 25$\%$ can be observed, the experimental spectra are  rather well reproduced. 
In particular, the shape of the p\pip\ invariant mass distribution around cos $\theta_n$=0 seems to indicate  that the \del\ 
contribution is still dominant in this  region.  The introduced flattening of the angular distribution of the $\Delta$ production  both compensates the missing yield around cos $\theta_n$=0 in the 
original model and gives better agreement of the invariant mass spectra. These distributions definitely contain  rich 
information about the pion production mechanism and should be compared to more sophisticated models including interference effects and non-resonant contribution.  Thanks to the high statistics, these detailed distributions  can indeed provide constraints  which are complementary to the results from the bubble chamber experiments \cite{Bugg64,Eisner65}.

\subsection{Analysis of the pp$\mathbf{\to}\mathbf{pp}\mathbf{\pi ^0} $ channel}
\label{sec|pizero}
We will now discuss the results obtained in the \pppiz\ channel and 
compare them to the resonance model in its standard and modified versions.
 
\subsubsection{Invariant masses}
 As mentioned in sec.~\ref{sec|delta}, model A, based mainly on the Teis resonance model \cite{Teis97},  underestimates the yield in the \pppiz\ channel by 20 $\%$  at 1.25 GeV and 35 $\%$ at 2.2 GeV (see fig.~\ref{fig:invariant_masses_small}). 
 After the inclusion of the changes in the  model B  motivated by the study of the \pppnpip\ channel (see sec.~\ref{sec|modif}), i.e. a slight rescaling of the cross sections of the different channels and the use of a phenomenological angular distribution for the \del\ production,  both the yields and the shapes of the \piz p invariant mass distribution are better reproduced, as demonstrated by the long dash-dotted curves in fig.~\ref{fig:invariant_masses_small}.
 
 At 1.25 GeV, 
  the change of the \del\ production angular distribution in the model mainly results  in a global increase of the cross sections in the HADES acceptance by 33$\%$,    with small effect on the shape of the \piz p invariant mass distribution. Note that, contrary to the \pnpip\ channel, no rescaling is applied to the model.  In contrast to  the pp$\to$pn\pip\ case, where the whole \del\ production angular distribution could be measured,  \del\ production in forward or backward angles is suppressed by the HADES acceptance in the pp$\to$pp\piz\ channel. This is the reason of the higher yield obtained for the simulation with model B, where the \del\ production angular distribution is flatter. \par
  At 2.2 GeV, the change of the shape of the \piz p invariant mass distribution is induced  by the increase of the cross sections for the higher lying resonances (N*(1440), N*(1520), N*(1535)) and the introduction of a non-resonant contribution. \par
   To summarize, the changes motivated by the study of the analysis of the \pppnpip\ channel also improve the description of the yields in the \pppppiz\ channel, which adds consistency to the procedure.  To complete the study, the proton  angular distributions measured at 1.25 GeV are investigated in the next section.

\subsubsection{Proton angular distributions at 1.25 GeV}
\begin{figure}
	\centering
		\resizebox{0.4\textwidth}{!}{\includegraphics{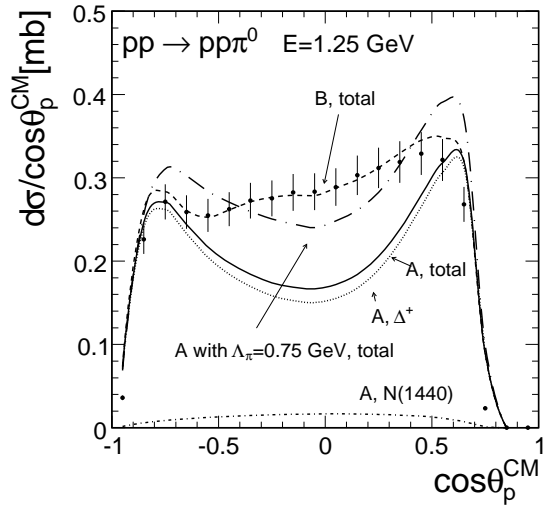}}
	\caption{Center-of-mass proton angular distribution  in the reaction \pppppiz\ at 1.25 GeV compared to three versions of the resonance model: model A (full curve), modified version of model A with $\Lambda_{\pi}$=0.75 GeV (long dot-dashed curve), and  model B (dashed curve). The \delp\ and N$^{\star}$(1440) contributions are depicted separately for model A by dotted and short dot-dashed curves, respectively.}
	\label{fig:pppi0_angp}
\end{figure}
 Even at 1.25 GeV, where pp$\to$p\del\ is the dominant process, the  \delp\ resonance cannot be unambiguously reconstructed due to the two  protons in the exit channel. 
However, although both, the proton coming from the decay of the \del\ resonance and the scattered one, contribute, the shape of their angular distribution inside the HADES acceptance is mainly sensitive to the distribution of \delp\ production angle and depends only marginally on the decay angle in our simple two-step model. It is therefore interesting to check whether the distribution of the proton angle allows  to draw   conclusions  on the distribution of the \del\ angle,  which are consistent with the pn\pip\ channel.
As can be seen in fig.~\ref{fig:pppi0_angp}, the acceptance  is limited to intermediate proton angles (about $ 45 ^{\circ} < \theta_p^{CM} < 155^{\circ}$ )  in the center-of-mass. The experimental distribution is however clearly much flatter than predicted by the simulation based on the  resonance model A (see sec.~\ref{sec|res_model}).  Changing the $\Lambda_{\pi}$ parameter in the vertex form factor from  0.63 to  0.75 GeV, as motivated by the  analysis of the \pppnpip\ reaction, the simulation comes closer to the data, although the yield around cos $\theta_p$=0 is still too low. This is related to the remaining underestimation of the $\cos\theta_n$ distribution in the pp$\to$pn\pip channel. A better agreement can indeed be obtained with the model B, which uses as an input for the \del\ production angular distribution the distribution fitted to  the measurement in the  \pppnpip\ channel (see sec.~\ref{sec|angn}), as shown by the dashed curve in the picture. 
This confirms that the two isospin channels can  be described consistently with the same \del\ production angular distribution. 
   Thus, model B can be exploited for the analysis of the exclusive pp$\to$pp\epem\ channel at 1.25 GeV, where a realistic model for the pp$\to$p\delp\ reaction is needed.

\subsection{Exclusive one-pion and one-eta production cross sections}
\label{sec|crossections}

\begin{figure}[h]
	\centering
\resizebox{0.5\textwidth}{!}{
		\includegraphics{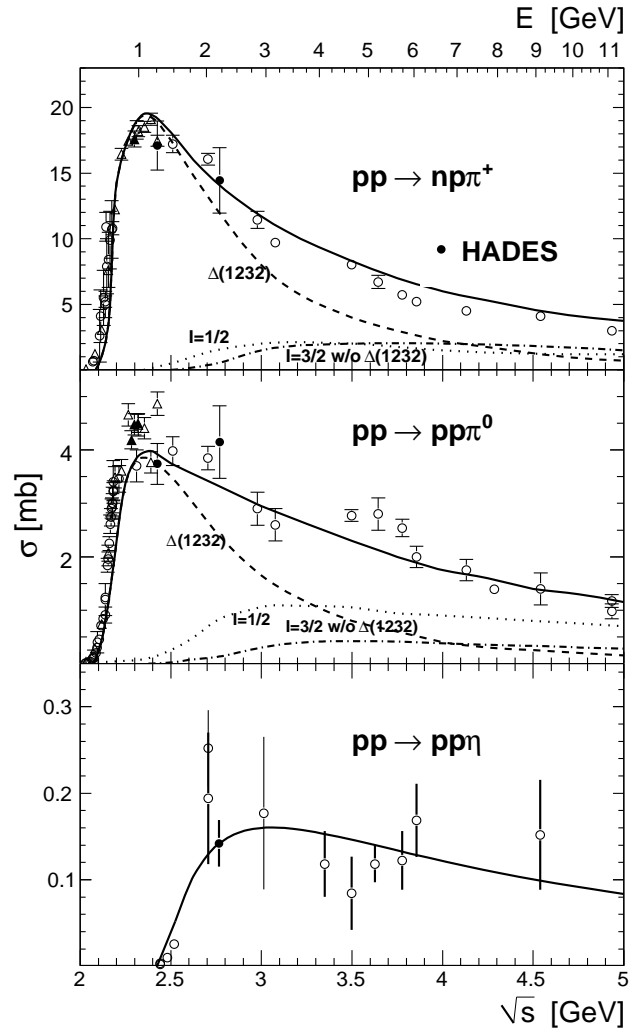}}
	\caption{Cross sections measured by HADES (full  dots) in hadronic channels for the \pppnpip (top), \pppppiz (middle) and pp$\to$pp$\eta$ (bottom)  reactions  compared to existing data (empty dots \cite{Baldini88,Chiavassa94,Betigeri02}, empty triangles  \cite{Shimizu82}, full triangles \cite{Sarantsev04,Ermakov11}).  The curves display the resonance model cross sections \cite{Teis97} (total: full curve, \del (1232): dashed curve, I=1/2: dotted curve, I=3/2 other than \del (1232): dashed-dotted curve).}
	\label{fig:sysHADES}
\end{figure}
The cross sections for the different one-meson production channels are reported in the first row of table~\ref{tab|cross_exp}. 
For the \pppnpip\ channel at 1.25 GeV,  the cross section was obtained by integrating over cos $\theta_n$ the acceptance corrected neutron angular distribution  (see sec.~\ref{sec|angn}).
 For the other channels, the  modified resonance model (model B) was used to extrapolate the  measured  yields   to 4$\pi$. More precisely, the acceptance factors were calculated as a ratio of the number of events generated in the simulation to the number of events after filtering by the HADES geometrical acceptance and analysis cuts. 
   To estimate the model dependence of these corrections,   parameters of the model were varied, especially the \del\ decay angular distribution for the pion production channels and the proportion of resonant contribution in the case of the $\eta$ production channel. The corresponding numbers are indicated in the second row of table~\ref{tab|cross_exp}. The main sources of errors are 
 the model dependence of the acceptance corrections (second row),   the normalization procedure (third row) and the efficiency corrections (fourth row). The errors due to the event selection, following the procedures discussed in sec.~\ref{sec|pionselect} and \ref{sec|etahad}  are also indicated (fifth row). Statistical errors (sixth row) are negligible. Note that the contribution of the error on the $\eta$ branching ratio into 3 pions is also of the order of 1$\%$. The obtained cross section values are compatible with previous measurements, as can be seen from 
fig.~\ref{fig:sysHADES}.\par
For the pion production channels, most of the data points obtained for $\sqrt
{s}$ between 2.0 and 2.4  GeV come from KEK \cite{Shimizu82} (black dots) and were not included in the Teis fits, which were based on CERN data tables \cite{Baldini88}. For the \pip\ production, the  KEK  points \cite{Shimizu82} fit rather well with the Teis curve as well as with  previous data \cite{Eisner65,Fickinger62}, while they are about  15-20$\%$ higher for the \piz\ production. 
Our data are compatible with the Teis curve, despite a slight underestimate of the \piz\ production at 2.2 GeV. 
As already mentioned, the relatively large error bars of our data with respect to the existing previous data are due to the combined effects of efficiency corrections, normalization and acceptance corrections, which were reduced in the case of bubble chamber experiments. The scattering of this data collection might however point  to an underestimate of   the error related to the event identification. 
From the HADES measurements,  the ratios of \pppnpip\ to \pppppiz\ cross sections  4.57$\pm$0.54 at 1.25 GeV and 3.49$\pm$0.63 at 2.2 GeV can be deduced, which has to be compared with the factor 5 expected in the case of \del\ excitation only, see eq.(\ref{eq|isospin}).  \par
As for the $\eta$ production, our experiment brings a new measurement ($\sigma =0.142 \pm 0.022$ mb) of the exclusive production cross section in pp reaction, in a region, about 230 MeV above the threshold,  where only the two  measurements from  Pickup {\it et al.} \cite{Pickup62} existed. Our point is in agreement with their value obtained in neutral channels, $\sigma=0.197 \pm 0.077$ mb, while  it is found  below their more precise measurement obtained in the three-pion  channel ($\sigma =0.242 \pm 0.043$ mb).  The quoted error might however be underestimated, as discussed in \cite{Balestra04},  considering the uncertainty due to the non resonant  background subtraction.   The cross section parameterizations used in \cite{Balestra04} and based on fits of data with $\sqrt{s}$ ranging from threshold up to 3.4 GeV did not take into account these points  and provided, at an energy of 2.2 GeV, values  between 70 and 100 $\mu b$, which are much lower than both Pickup's results \cite{Pickup62} and the value measured in the present experiment.  Our measurement is in very good agreement with the resonance model \cite{Teis97}, where the $\eta$'s  are assumed to be produced only via  N(1535) resonance decay. This assumption of fully resonant production seems however in contradiction with the DISTO analysis \cite{Balestra04}.   Our new measurement can hence be used to test various  models of $\eta$ production. Previous OBE calculations \cite{Laget91,Vetter91,Kaptari10}  showed deviations of a factor 2 depending on the values of the $\rho$NN$^{\star}$(1535) and  $\omega$NN$^{\star}$(1535) coupling constants, which should be updated in view of the actual constraints on these parameters,  as was done recently closer to threshold \cite{Nakayama08}. 
%

\begin{table}[h!]
\centering
\begin{scriptsize}
\begin{tabular}{|c|c|c|c|c|c|}
\hline
reaction & \multicolumn{2}{|c|}{\pppnpip}  & \multicolumn{2}{|c|}{\pppppiz} & \ppppeta  \\ \hline
\hline
energy & 1.25 GeV&2.2 GeV  & 1.25 GeV & 2.2 GeV& 2.2 GeV  \\ \hline
cross section  (mb) & 17.1 $\pm$ 2.0  & 
                14.45 $\pm$ 3.2 & 
                 3.74 $\pm$ 0.48  & 
                4.15 $\pm$ 0.85 & 
                0.142 $\pm$ 0.022 \\ \hline
acceptance corrections &$\pm$ 1.0 & $\pm$1.1 &$\pm$ 0.2 & $\pm$0.2 &$\pm$ 0.006\\\hline                
normalization & $\pm$ 1.1 & $\pm$ 1.6  & $\pm$ 0.25  & $\pm$ 0.46 &  $\pm$ 0.016   \\ \hline                
efficiency &$\pm$ 1.3 & $\pm$2.5 & $\pm$0.33 &$\pm$ 0.65 & $\pm$0.013\\ \hline   
event selection& $\pm$ 0.3 & $\pm$ 0.7 & $\pm$ 0.12 & $\pm$  0.2 &$\pm$  0.005 \\ \hline               
statistics & $\pm$ 0.01 & $\pm$ 0.01 & $\pm$  0.003  & $\pm$ 0.004 & $\pm$ 0.002 \\ \hline                
\end{tabular}
\end{scriptsize}
\caption{Cross sections for exclusive meson production channels measured by HADES in hadronic channels are given with the total error, calculated as the quadratic sum of  the statistic and systematic errors listed in the following rows.}
\label{tab|cross_exp} 
\end{table}

\subsection{$\mathbf{\pi^0}$ and $\mathbf{\eta}$ Dalitz decay analysis}
\label{sec|piz_eta_results}
\subsubsection{Dielectron invariant mass}
\begin{figure}
\begin{center}
  \resizebox{0.48\textwidth}{!}{	  
     \includegraphics{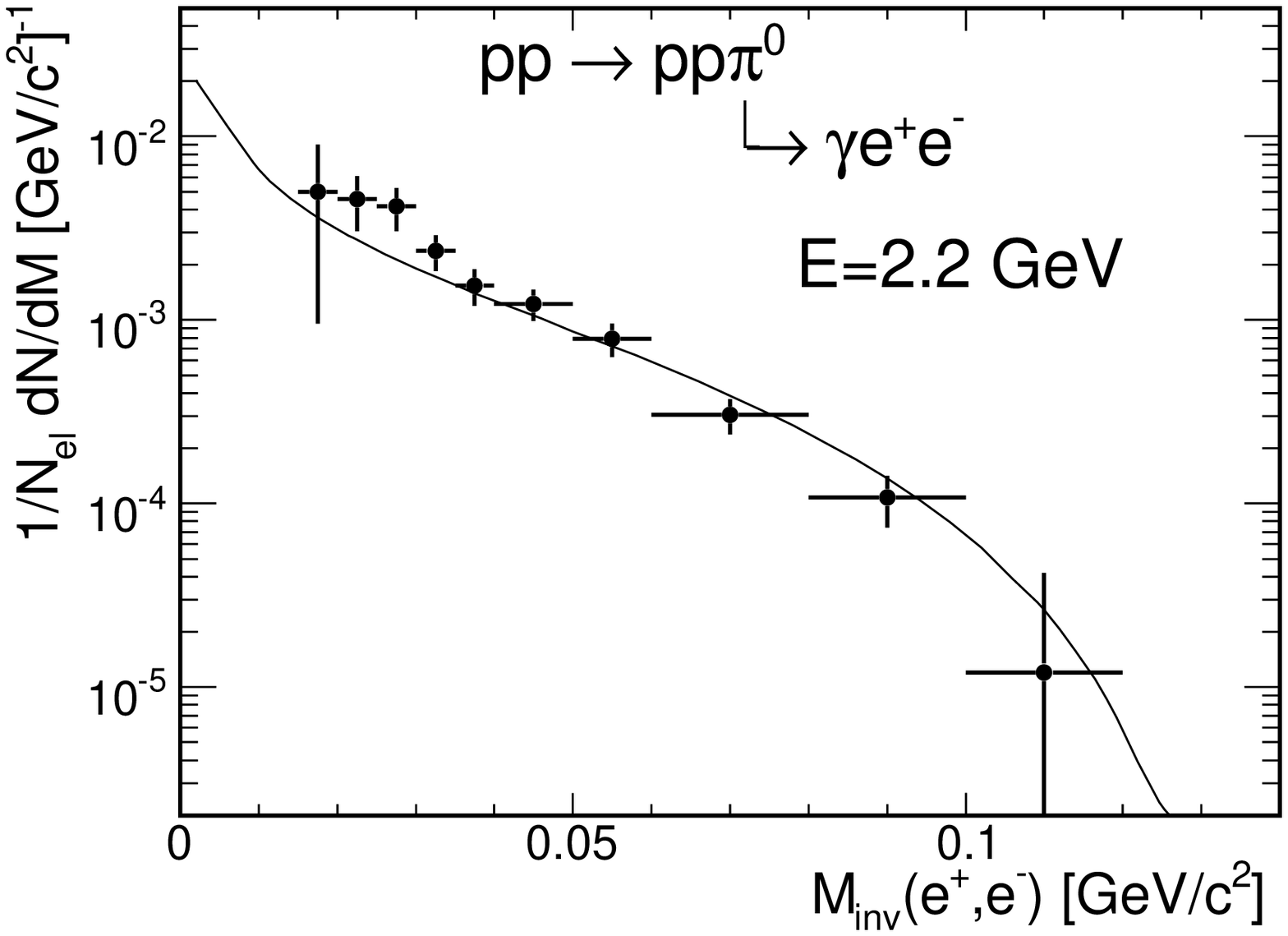}}
      \resizebox{0.48\textwidth}{!}{	  
		\includegraphics{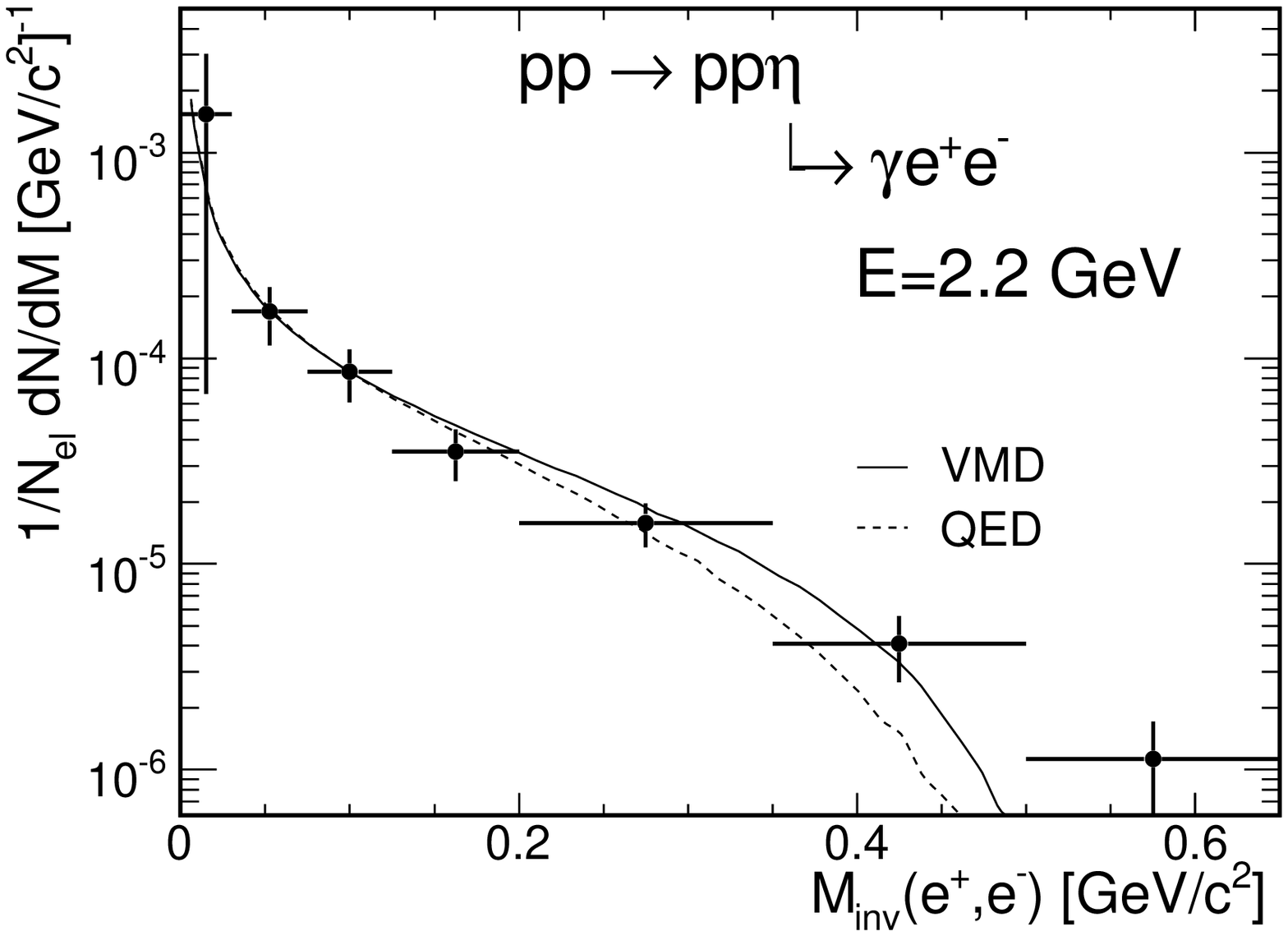}}
         \caption{\small Distribution of di-electron invariant
          masses for the \piz\ (left) and $\eta$ (right) Dalitz decays, obtained after efficiency and acceptance corrections and compared to the simulations with model A (full line), including VMD models. The yields have been divided by the elastic scattering yields. In the case of the $\eta$, the  simulation without $\eta$ form factor (labeled as QED) is shown as a dashed curve for comparison.   
}
          \label{minv}	
          
\end{center}
\end{figure}

As already explained in sec.~\ref{sec|pi0etadal_signal}, the $\pi^0$ and $\eta$ Dalitz decay signals  has been extracted, in each \epem\ invariant mass bin, using the  missing masses to the pp and pp\epem\ systems and were then efficiency corrected. In addition, an acceptance correction obtained from simulations with the resonance model was applied.
The resulting \epem\ invariant mass distributions are displayed in fig.~\ref{minv}, with statistical errors and systematic errors added quadratically. In  the case of the \piz , the largest source of systematic error is  the rejection of \epem\ pairs from photon conversion, while in the $\eta$ region,  it is due to \epem\ pairs from \piz\ decay in multipion production processes.\par
 The experimental values are compared to the results of the simulation, with exclusive meson production cross sections from table~\ref{tab|res_channels} and branching ratios  from table~\ref{tab|exit_channels}. 
The good agreement obtained for both the \piz\ and $\eta$  peak  is therefore a check of the consistent extraction of the corresponding signals, which is very useful for all dielectron analyses performed with the HADES detector. The small excess around 0.03 \gcsq\ is most likely due to a remaining   contamination of conversion pairs. The possible  contribution of Dalitz decays of baryon resonances, corresponding to a pp\epem\ final state, has also been investigated and is found to be negligible, except in the mass region  close to the kinematical limit (M$_{inv}$(e$^+$,e$^-$) = 0.547 \gcsq ), which could possibly explain that the measured yield  for the reaction \ppppeta\ is higher than the simulation above 0.5 \gcsq . \par
It has been checked that the  shapes of these spectra do not depend on the ingredients of the simulation related to the meson production mechanisms, like the relative yields of the different resonant contributions, but are characteristic of their Dalitz decay. 
The description of these Dalitz decay processes in the simulation implies electromagnetic form factors which can be  implemented in the simulation following the VMD model \cite{Froehlich07,Landsberg85}.  Modelling the transitions  as point-like (refered to as QED) or using VMD form factors lead to negligible differences   for the \piztodal\ case  and show up for the  \etatodal\ case only at larger values of the \epem\ invariant mass.  There, our data are however  not precise enough to provide any further quantitative constraint to these models. \par 
The yields are well reproduced by the  simulation with meson production cross sections from table \ref{tab|res_channels}.  
 In the case of the $\eta$ production, since the  cross section is only fixed by our measurement in the hadronic channel   (sec.~\ref{sec|crossections}), this shows the consistency of the hadronic and leptonic reconstructions and the good control of the corresponding efficiencies.  More quantitatively, the ratio of yields measured in  \pip \pim \piz\ and  $\gamma$\epem\ decays of the $\eta$ meson is 218$\pm$25, i.e. fully consistent with  the value 230$\pm$5 given by  the simulation.
   In the case of the \piz , where the production cross sections are constrained by independent data, the analysis of the \piztodal\ channel  provides  a global consistency check of the whole analysis chain for dileptons. 
 
\subsubsection{Helicity angle}

\begin{figure} 
\begin{center}
        \resizebox{0.48\textwidth}{!}  {\includegraphics{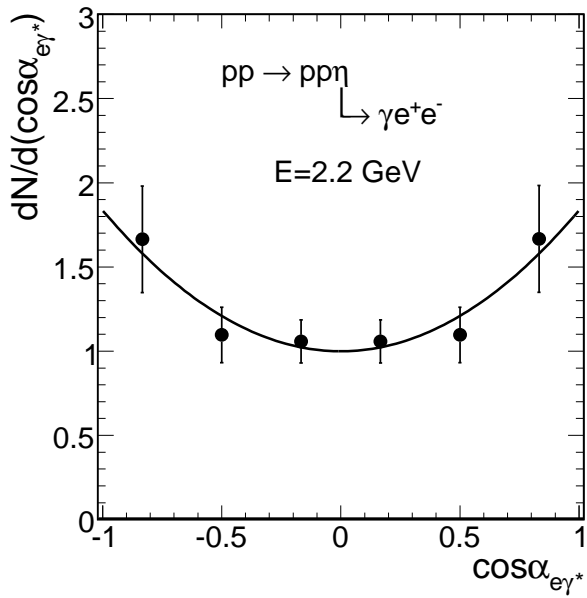}}
              \caption{\small Angular distributions of electrons and positrons in the virtual photon rest frame (helicity) after acceptance and efficiency correction, using an arbitrary normalization. The solid curve is the fit using (a(1+bcos$^2 \alpha_{e\gamma^{\star}})$) with b=0.98$\pm$0.48.}
   \label{fig|helicity}
\end{center}
\end{figure}
An interesting feature of the Dalitz decay of pseudo-scalar mesons is the  polarization of the virtual photon which is transverse. As a consequence, the distribution of the helicity angle $\alpha_{e\gamma^{\star}}$  follows  1+cos$^{2}\alpha_{e\gamma^{\star}}$.
 The calculation of this angle first implies a boost of all particles in the meson rest frame. Then the helicity angle is defined as the 
polar angle of  the electrons in  the virtual photon rest frame, with respect to the virtual photon direction. 
The acceptance and efficiency corrections were calculated using the simulation of the dielectron production via $\eta$ Dalitz decay, as described above. As shown in fig.~\ref{fig|helicity},  this angular distribution can be fitted  by a function of the form  a(1+bcos$^2 \alpha_{e\gamma^{\star}})$  with
$b=0.98 \pm 0.48$,  in agreement with the QED prediction,  $b$=1 \cite{Bratko95}.  
\par 
  With the HADES set-up, it is therefore possible to reconstruct the helicity angle distribution
of the $\eta$  Dalitz decays. The extraction of the anisotropy parameter $b$ should then  be also possible in the case of the \del\ Dalitz decay, where, the  polarization of the virtual photon is also mainly transverse, since the Coulomb amplitude in the N\del\ transition is small and hence a 1+cos$^2 \alpha_{e\gamma^{\star}}$ distribution is expected.  Helicity angle distributions have also been investigated 
in heavy-ion reactions \cite{Agakishiev_ArKCl2011} in order to identify the nature of the ``excess'' beyond the $\eta$ contribution. 

\section{Summary and outlook}
\label{sec|conclusion}
HADES has provided a measurement of the reactions \pppnpip\ and   \pppppiz\ at
1.25 GeV and 2.2 GeV and \ppppeta\ at 2.2 GeV  using both hadronic and leptonic channels.  Using the hadronic channels, high statistics differential cross sections could be measured in the HADES acceptance. In addition,  integrated cross sections were extracted for all these channels and the neutron angular distribution in the \pppnpip\ reaction at 1.25 GeV was fully reconstructed.  These data allow to test pion production mechanisms and the contribution of baryonic resonances with a high statistical precision, in complement to previous low-statistics but high-acceptance experiments.   
We left for further studies the comparison of these data  to calculations including resonant and  non resonant contributions in a coherent way. Our aim in this paper was  twofold: first, to show  the sensitivity of the present data to the ingredients of the transport models used for the dielectron production, which are based on resonance models and, second, to obtain a parameterization of meson and baryon resonance production  for  dielectron channels analysis. Following this line, an analysis based on a resonance model \cite{Teis97} was presented.  
An overall agreement with the original model is shown, but a better description  could be obtained in both isospin channels, using at 1.25 GeV  a less steep angular distribution for the \del (1232) resonance production and at 2.2 GeV an increased production cross section  for the higher lying resonances.  A precise description of the \del (1232)  production angular distribution at 1.25 GeV is especially important for the on-going analysis of the Dalitz decay of the \del (1232) resonance using the pp$\to$pp\epem channel.  On the other hand, further information on higher lying resonances can be gained by studying two-pion production channels, which were also recently measured in the HADES experiments.  
The present determination of the exclusive $\eta$ production cross section is most important, as it provides the first precise measurement of the exclusive production cross section in a region where deviating   model predictions can be found.  
\par 
      The reconstruction of \piz\ and $\eta$ Dalitz decay signals presented in this paper is fully consistent with the hadronic channels, and the invariant masses and acceptance corrected  helicity angle distributions are in good agreement with QED predictions. These results confirm the ability of HADES to reconstruct sensitive observables in dielectron channels, which is a very important consistency check for previous and next-coming analyses. The helicity angle was  used to study dielectron sources in heavy-ion reactions \cite{Agakishiev_ArKCl2011} in different invariant mass regions and is also  used to discriminate the \del (1232) Dalitz decay process from the pp Bremsstrahlung contribution in the on-going analysis of the exclusive pp\epem\ channel in pp reactions at 1.25 GeV \cite{Ramstein10}. \par
 As a final conclusion, the present analysis provided  important consistency checks for  dielectron studies, as well as precise results for meson production measured in hadronic channels, paving the way for further theoretical or experimental studies of exclusive  dielectron and hadronic channels in elementary reactions.\par
\section{Acknowledgements}
The collaboration gratefully acknowledges the support by
CNRS/IN2P3 and IPN Orsay (France), by SIP JUC Cracow (Poland) (NN202 286038, NN202198639), by HZDR, Dresden (Germany) (BMBF 06DR9059D), by TU M\"{u}nchen, Garching (Germany) (MLL M\"{u}nchen, DFG EClust 153, VH-NG-330, BMBF 06MT9156 TP5, GSI TMKrue 1012), by Goethe-University, Frankfurt (Germany) (HA216/EMMI, HIC for FAIR (LOEWE), BMBF 06FY9100I, GSI F$\&$E), by INFN (Italy), by NPI AS CR, Rez (Czech Republic) (MSMT LC07050, GAASCR IAA100480803), by USC - Santiago  de Compostela (Spain) (CPAN:CSD2007-00042).

%

\end{document}